# Extended Pacejka Tire Model for Enhanced Vehicle Stability Control

Kanwar Bharat Singh* & Srikanth Sivaramakrishnan
Tire Vehicle Mechanics, The Goodyear Tire & Rubber Company

**ABSTRACT:** Despite their widespread use, current tire models have demonstrated a certain level of inaccuracy, primarily due to uncertainties related to unaccounted nonlinearity and disturbance effects resulting from tire operating conditions. Noteworthy factors such as tread depth, inflation pressure, tire temperature, and road surface condition significantly impact tire force and moment characteristics. These factors can vary considerably during tire operation and significantly affect both tire and vehicle performance. The improvement of tire models is crucial to enhance the effectiveness of advanced vehicle control systems, as accurate tire force characteristics are required for maintaining vehicle stability during demanding maneuvers. This paper investigates the impact of varying tire temperature, inflation pressure, and tread depth on steady-state tire force characteristics by analyzing the coefficients of the Pacejka 'magic formula' (MF) tire model. Based on this analysis, adaptation equations are proposed to compensate for the influence of these variables on the tire force curve. The advantages of using an adapted tire model are then demonstrated through simulation studies of a classical vehicle stability control system that can adapt to diverse operating conditions. A comparison is made between the adapted tire model-based controller and a controller based on a fixed reference model.

**Keywords:** tire model adaptation, magic formula, cornering stiffness, grip level, stability control

## 1. Introduction

Vehicle dynamic performance is predominantly controlled by tire dynamic characteristics through the forces and moments generated at the tire-road contact patch. Tire forces are a combination of two factors: friction/sliding in the contact patch and elastic deformation/slipping of the tire. These primary forces affecting planar vehicle motions are constrained based on the friction circle, as shown in Figure 1, where the tire is always subjected to a combination of turning and acceleration/braking forces.

*Corresponding author email: kanwar-bharat_singh@goodyear.com
Avenue Gordon Smith, 7750 Colmar-Berg, Luxembourg; +352 81 991

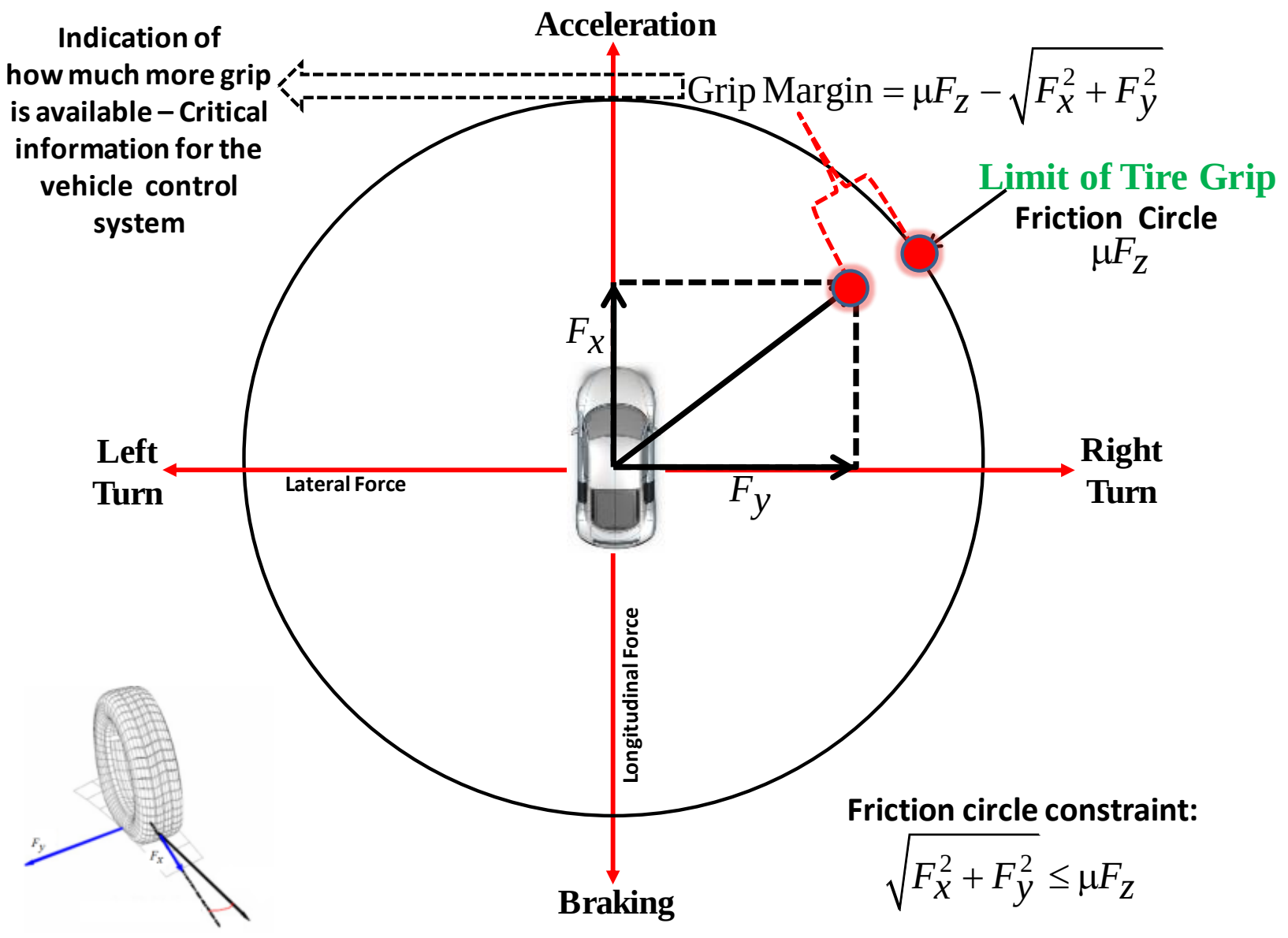

**Figure 1: Tire friction circle - resultant force and tire force reserve (grip margin)**

The control of the tire lateral forces benefits vehicle stability and handling characteristics. Thus, tire saturation constraints must be taken into account for a proper control system design, as this provides an indication of the maximum force that can be generated by the tire. For those control systems based on the simplified linear vehicle model, which ignores the nonlinear characteristic of the tire, the robustness subject to parameter uncertainties during vehicle driving cannot be guaranteed. Hence, the ability to reliably estimate tire forces is mainly control oriented and is functional to the implementing and maximizing the performance of vehicle control systems, which work well only when the tire force command computed by the safety systems is within the friction limit.

Electronic stability control (ESC) systems used in current production vehicles use feedback control to track a reference yaw rate using individual braking via direct yaw control (DYC), and/or an active front steering (AFS)/active rear steering (ARS) system. The ESC system is activated when the deviation between the reference and actual measured yaw-rate is larger than a preset dead-band. The reference yaw-rate is calculated based on a vehicle model using vehicle speed and steering wheel angle signal as inputs. The desired yaw-rate is calculated for tracking the driver’s intention. The reference vehicle model in the control system is based on tire characteristics (the front and rear cornering stiffness). Hence, the cornering stiffness of the tire is an important parameter for the controller. However, it fluctuates under varying road conditions, or under varying operating conditions of the tire, which affects the accuracy of the vehicle stability control. For example, inflation pressure increases as the contained air temperature increases with tire use. This pressure “build-up” necessarily changes tire force and moment characteristics and must be accounted for when selecting initial or “cold” inflation pressures. Tire tread depth is known to significantly change the tire cornering and braking capabilities. As the temperature changes, the modulus of elasticity will change to influence the stiffness and the coefficient of friction of the tire. For these reasons, it is highly desirable to have inflation pressure, tread depth, tire temperature, and road surface condition as an input to a tire model along with the traditional operating condition inputs (normal load, slip state, camber angle, vehicle speed, etc.).

Thus, to account for vehicle behavior in changing operating and environmental conditions (changing tire operating and road surface conditions) a further step is required, that of adapting the tire model to changing inflation pressure, temperature, tread depth, and road surface conditions. The availability of a high-fidelity tire model would facilitate the online computation of the optimized active longitudinal and lateral tire forces to achieve vehicle stability and safety without degrading driver intentions. This paper is the first step in addressing the technical challenges associated with adapting the Pacejka magic formula tire model [1] to tire inflation pressure and tread depth. The rationale behind the selection of temperature, inflation pressure and tread depth being the availability of these signals from tire attached sensor systems in the near future.

The purpose of this paper can be divided into two parts. The first part describes how tire temperature, inflation pressure and tread depth affect tire performance in terms of lateral force. The second part describes the modeling activities to account for such effects followed by the discussion of simulation results of a reference model-adaptive vehicle stability control to quantify the performance benefits over a controller utilizing a fixed reference model. The following section describes previously published work in this direction.

## 2. State-of- the Art Literature Review

Over the years various developments have been made to extend tire models to improve their prediction capabilities under various operating conditions. This section presents a descriptive overview of the State-of-the-Art of the work done towards the development of tire models specifically adapted to signals which are expected to be available from tire sensor systems in the near future.

### 2.a. Enhanced Magic Formula Tire Model to Handle Inflation Pressure Changes - Eindhoven University of Technology / TNO Automotive

As part of this study, the influence of inflation pressure on the tire characteristics was investigated and the magic formula extended to incorporate these effects [2]. This was achieved by adapting the tire cornering stiffness (Cy) and peak lateral friction coefficient ($\mu_{max}$) through scaling factors which have a quadratic relationship to pressure change as shown in Equations (1) and (2), with the fit parameters evaluated through a polynomial fit over experimental data. The expression for the variation of peak lateral grip with inflation pressure was suggested although no clear trends were observed from experimental data for various tires.

$$\mu_x = (p_{Dx1} + \mathrm{p}_{Dx2} df_z)(1 - p_{Dx3}\gamma^2)(1 + p_{x3} dp_i + p_{px4} dp_i^2)\lambda_{\mu x} \tag{1}$$

$$K_{x\kappa} = (p_{Kx1} + p_{Kx2} df_z)\exp(p_{Kx3} df_z)(1 + p_{px1} dp_i + p_{px2} {dp_i}^2)F_z\lambda_{Kx\kappa} \tag{2}$$

### 2.b. TaMe Tire Model - Michelin

The TaMeTire model is a thermo-mechanical tire model, developed and patented by Michelin [3-4] which estimates the forces and moments generated by a tire in real-time while accounting for variations in rubber properties due to temperature distribution in the tire. This model consists of three different interacting sub-models: a mechanical model, a local thermal model for estimating the variation of rubber properties such as shear modulus, rigidity and friction coefficient as a function of the tire surface temperature, and a global thermal model for predicting the temperature distribution along the thickness of the tread, as shown in Figure 2.

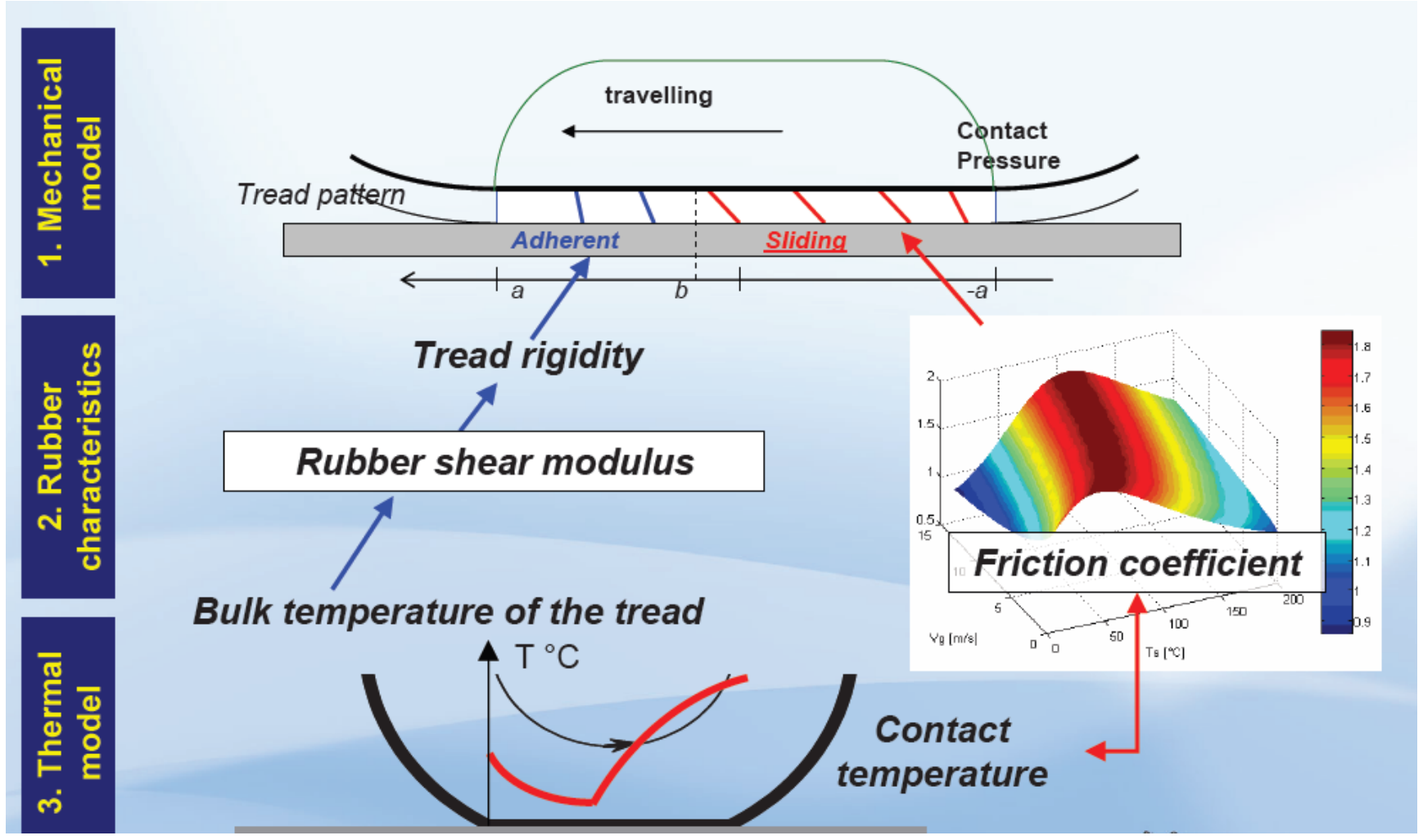

**Figure 2: Michelin TaMe tire model – modeling approach [4]**

The tire bulk temperature is calculated from the tire surface temperature using the first principle of thermodynamics with a set of assumed boundary conditions. This way of modeling the internal tire temperature distribution along with a detailed mechanical model show high prediction accuracy for a large range of sideslip angles and in all cases of tire inflation pressure, speed, normal load and camber angle, an example of which is shown in [Figure 3]. One can observe change in lateral force from the increasing to the decreasing phase of slip angle which can be attributed to the rise in surface temperature.

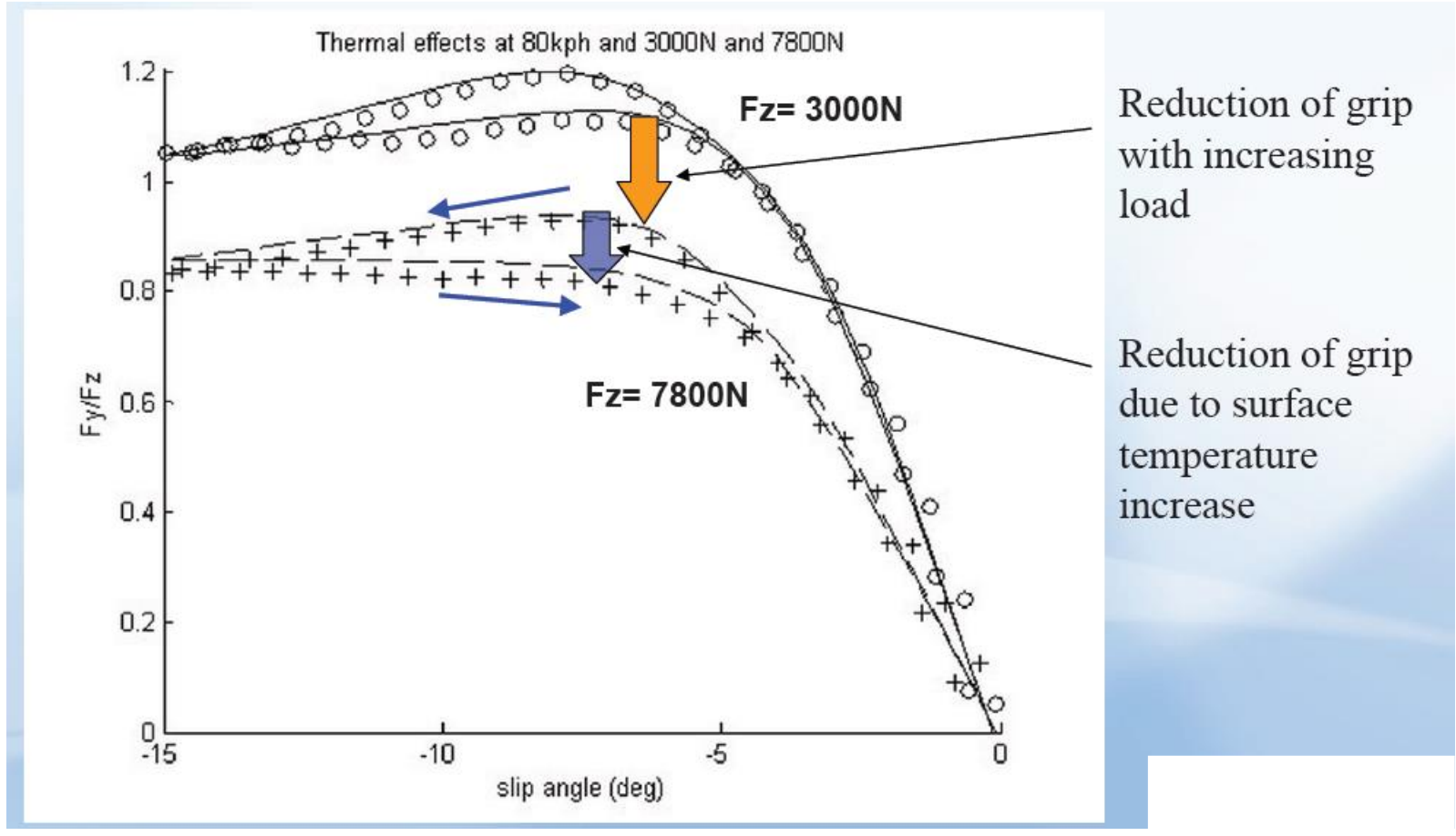

**Figure 3: Michelin TaMe tire model - lateral forces measured and modeled [4]**

### 2.c. Tire Lateral Force Model Based on the Pacejka Magic Formula with Influence of the Tire Surface Temperature - Toyota

This model developed by Toyota [6] is based on the magic formula and concentrates only on the prediction of lateral tire forces. In this, the surface temperature of the tire is calculated based on an energy balance equation at the contact patch.

$$\frac{dT}{dt} = \frac{1}{W}\left[F_y V\alpha - \lambda A(T - T_0)\right] \tag{3}$$

where

$W$ - Heat Capacity

$\lambda$ – Thermal conductivity

$A$ - Contact area

$T_0$ – Ambient temperature

$T$ - Surface temperature

$F_y$ - Tire side force

$\alpha$ – slip angle

$V$ - Vehicle velocity

The Pacejka coefficients were then scaled as a linear function of the tire surface temperature (T) with the derivative of the change of peak grip and cornering stiffness with temperature assumed to be a constant parameter. The modified Pacejka coefficients were now used to calculate the lateral force compensated for temperature. The prediction model was then validated with experimental data on a Flat-Trac machine.

$$F_y(\alpha, T) = D_y(T)\sin[C_y tan^{-1}\{B_y(T)\alpha - E_y(B_t(T)\alpha - tan^{-1}(B_y(T)\alpha))\}] + S_{vy}$$

$$\begin{aligned} D_y(T) &= D_{y0} \cdot \{1 + \frac{\partial \mu}{\partial T}(T - T_m)\} \\ K_y(T) &= K_{y0} \cdot \{1 + \frac{\partial C_p}{\partial T}(T - T_m)\} \\ B_y(T) &= \frac{K_y(T)}{C_y D_y(T)} \end{aligned} \tag{4}$$

where

$D_y(T)$ - Peak value of tire side force including the influence of the tire surface temperature

$D_{y0}$ - Peak value of steady state tire side force model

$K_y(T)$ - Cornering power including the influence of tire surface temperature

$K_{y0}$ - Cornering power of the steady state tire model

$T_m$ - Average tire temperature during measurement to obtain the steady state tire model data

$\frac{\partial C_p}{\partial T}$ – Change in tire side force with tire surface temperature

$\frac{\partial \mu}{\partial T}$ – Change in cornering power with tire surface temperature

### 2.d. Enhanced Tire Lateral Force Model with Influence of the Tire Surface Temperature - Chalmers University of Technology, Sweden and Renault SA

The main objective of this study by Renault [5] involved the analysis of all factors influencing tire surface temperature and to consequently develop a thermal tire model that can predict lateral forces including the influence of surface temperature. This was achieved by developing a quadratic function based empirical model that predicts the temperature at the surface of the tire based on slip angle, camber angle, vehicle speed, inflation pressure and vertical load. The parameters of the quadratic function were estimated through an optimization routine based on Flat-Trac® experimental data.

$$T(t=n) = C_k T(t=n-1) + \sum_{i=1}^{5} C_k\, P_i^2 + \sum_{i=1}^{5} C_k\, P_i + \sum_{i=1}^{5} \sum_{j=2}^{5} C_k\, P_i P_j + C_k \tag{5}$$

Where

$C_k$ - Constant coefficients determined using indoor experiments

$P_i$ – Input parameters

This predicted value is then used to scale magic formula coefficients according to the Toyota tire model [6] to estimate lateral forces. This model was later modified for validation on the track by including the effect of wheelhouse ambient temperature and loss due to forced-air convection. Comparison with experimental results, as shown in Figure 4 indicate that the thermal tire model can predict both tire surface temperature and lateral forces at higher temperatures.

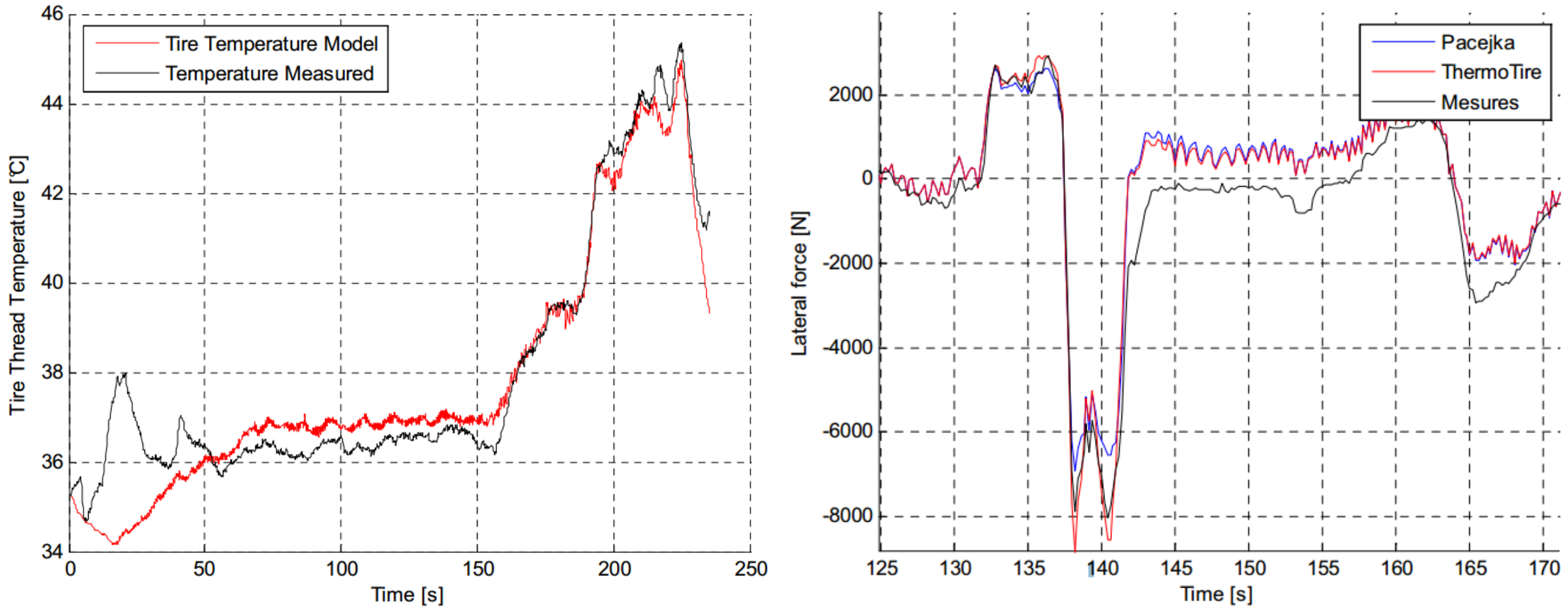

**Figure 4: Prediction of surface temperature and lateral forces [5]**

### 2.e. Modified Brush Model with Temperature Effects for Lap Time Simulations for Combined Slip Maneuvers – University of Surrey and Politecnico di Torino

This model was developed at the University of Surrey [7] because of a study aimed towards specifically developing a thermal tire model for racing applications. In this, both lateral and longitudinal stiffness in addition to peak grip is assumed to linearly increase with tread temperature. This is used in the brush model to calculate both lateral and longitudinal forces.

$$\begin{aligned} C_{pX,Y} &= C_{pX,Y,0} + k_{pX,Y}(T - T_0) \\ \mu_{pX,Y} &= \mu_{X_0,Y_0} + k_{\mu X,Y}(T - T_0) \end{aligned} \tag{6}$$

The tread temperature is calculated using two differential equations which model the heat transfer in the carcass and the tread as shown in equations below.

$$C_{eq_carcass}\frac{dT_{carcass}}{dt} = P_{rolling_resistance} + P_{conduction} + P_{ambient,carcass}$$
$$C_{eq_tread}\frac{dT_{tread}}{dt} = P_{F_x,tire} + P_{F_y,tire} - P_{conduction} + P_{ambient,tread} \quad (7)$$

The model schematic is illustrated in Figure 5

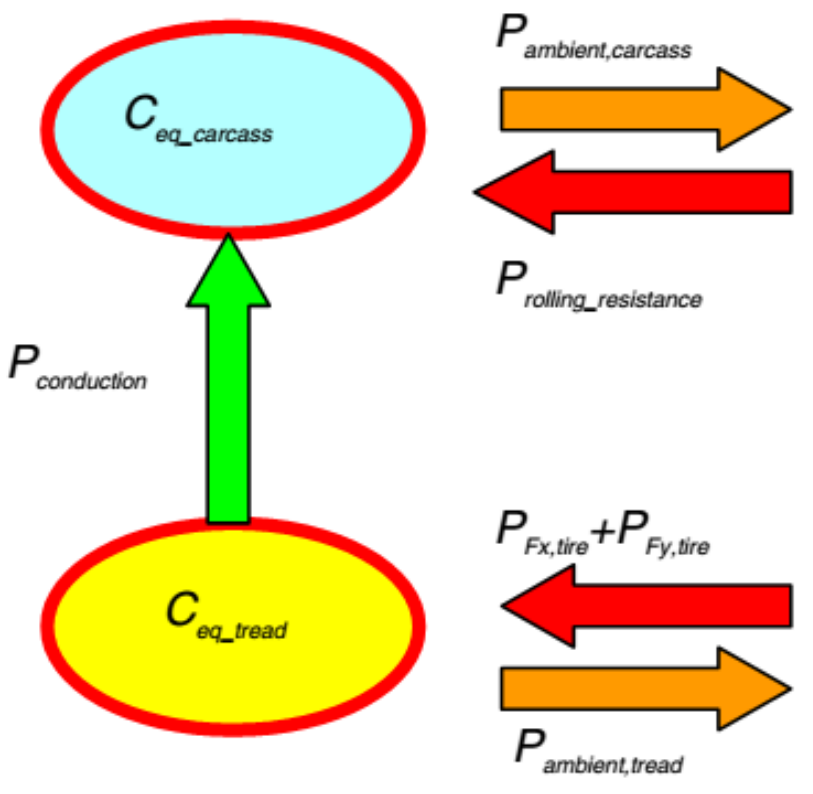

**Figure 5: Thermal model schematic [7]**

The experimental validation of the model under combined slip maneuvers as shown in Figure 6 indicates that the modified brush model adapted to temperature can provide a better prediction of both lateral and longitudinal forces.

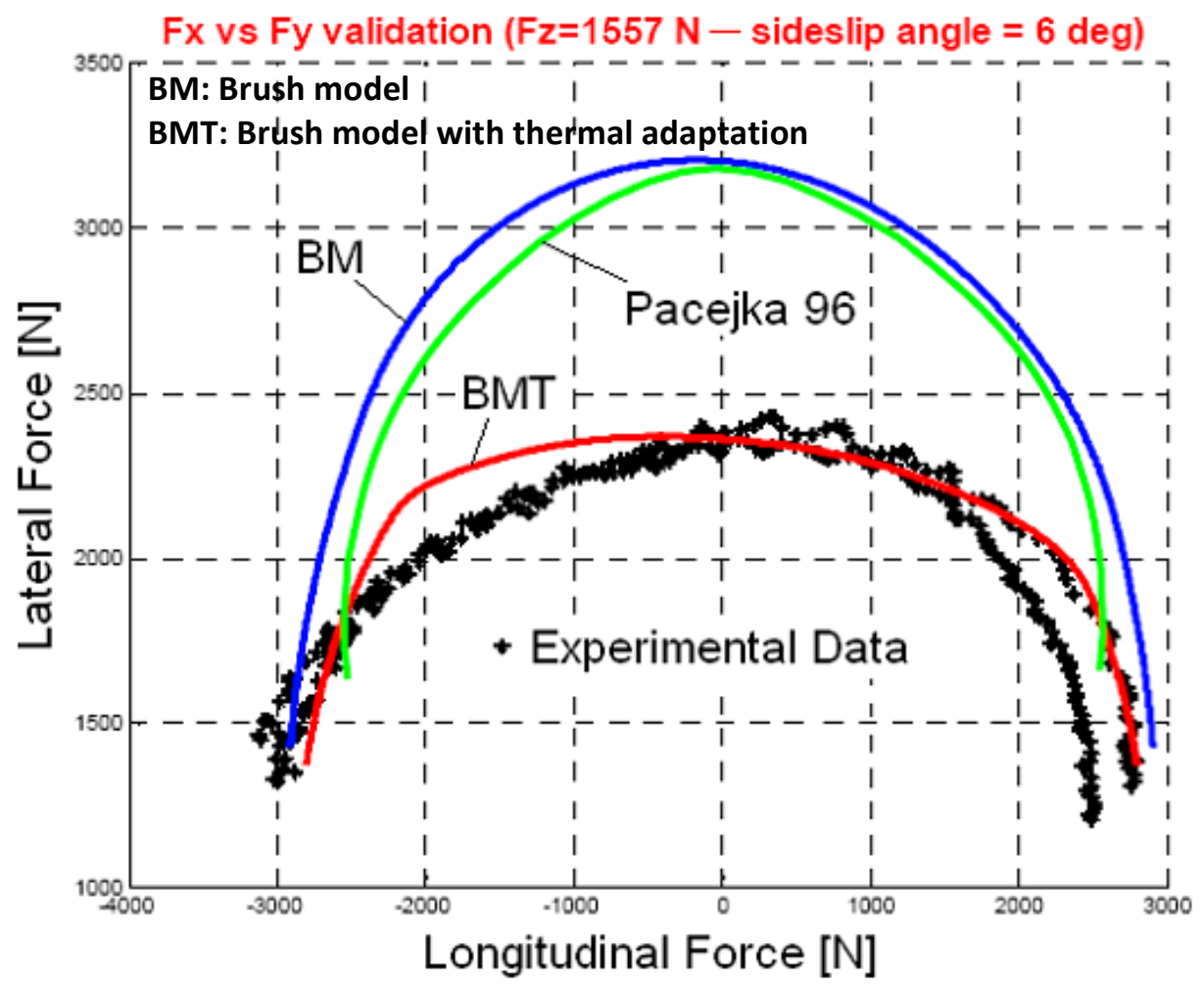

**Figure 6: Experimental validation of the modified brush model [7]**

A summary of the state-of-the-art literature review is given in Table 1.

**Table 1: State-of-the-Art literature review – summary**

| Organization | Adaptation Variable | Model Input(s) | Reference |
|---|---|---|---|
| Eindhoven University of Technology / TNO Automotive | MF - Pressure Adaptation | Cavity Pressure | [2] |
| Michelin | TaMe Tire Model - Temperature Adaptation | Tire Surface and Internal (Bulk) Temperature, Track Temperature, Ambient Temperature | [3-4] |
| Chalmers University of Technology, Sweden and Renault SA | Tire Lateral Force Model -Temperature Adaptation | Tire Surface Temperature | [5] |
| Toyota | Tire Lateral Force Model -Temperature Adaptation | Tire Surface Temperature | [6] |
| University of Surrey and Politecnico di Torino | Combined slip brush model – temperature adaptation | Tire Surface Temperature | [7] |

## 3. Quantifying the Influence of Variations in the Tire Inflation Pressure, Tread Depth, load and temperature on the Steady State Tire Force Characteristics

In its simplest formulation, a typical tire model [19] describes the relationship between the tire force and the slip as a function of two parameters, namely, cornering stiffness and the peak friction force coefficient [Figure 7], and hence these two parameters were selected as the parameters of interest for this study

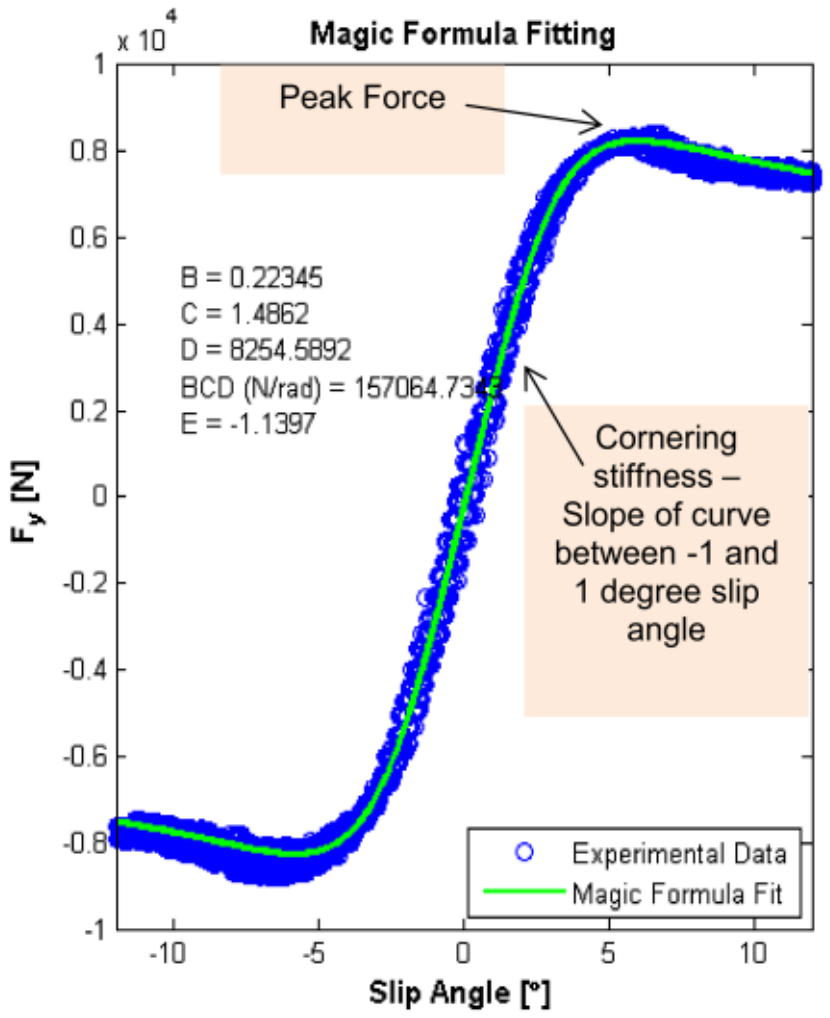

**Figure 7: Tire parameters of interest**

With the objective of quantifying the influence of variations in the tire inflation pressure, tread depth, normal load, and temperature on the tire cornering stiffness and the peak friction force coefficient, experiments were conducted on the Flat-Trac® machine [Figure 8].

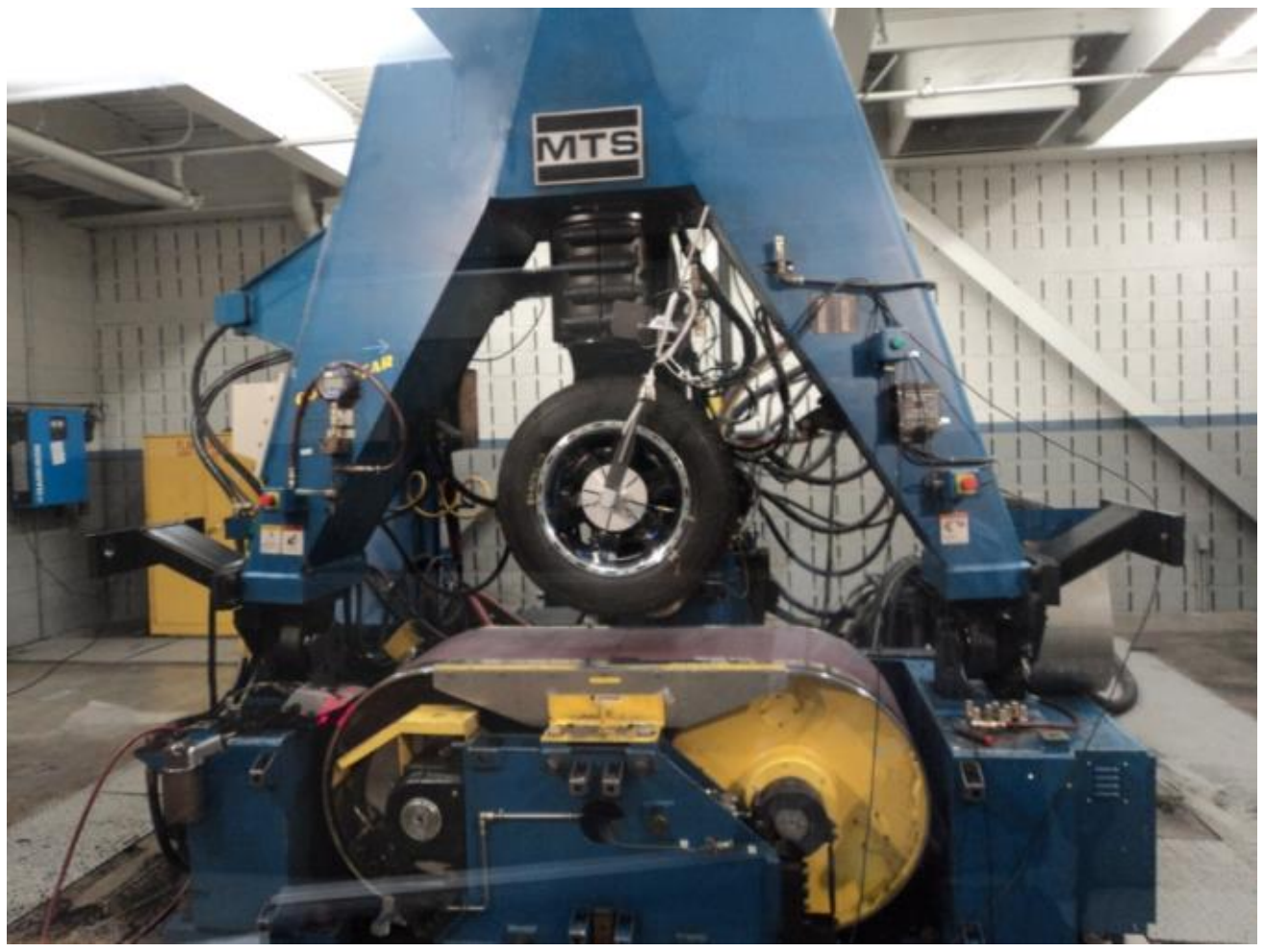

**Figure 8: Flat-Trac® machine**

The sensitivity study was conducted for a high-performance summer tire. In addition to this, the influence of different constructions such as summer and all-season tires were also analyzed as vehicles in European countries typically have different tires mounted based on the season (summer or winter), mainly due to government mandated requirements. In the first round of tests, the influence of temperature was eliminated by keeping the slip angle sweep rate and the tire speed constant [Table 2].

**Table 2: Variable dependency chart**

| | Tire Surface Temperature | Tire Bulk Temperature | Inflation Pressure | Normal Load | Speed | Tread Depth |
|---|---|---|---|---|---|---|
| Cornering Stiffness | x (~Constant) | x (~Constant) | ✓ | ✓ | x (Constant) | ✓ |
| Peak Friction Coefficient | x (~Constant) | x (~Constant) | ✓ | ✓ | x (Constant) | ✓ |

✓: varied
x: kept constant

- To evaluate the influence of the inflation pressure on the tire characteristics, four levels of pressure were analyzed: (a) 33psi, (b) 37psi, (c) 41psi, (d) 45psi.
- To evaluate the influence of the tire tread depth on the tire characteristics, three levels of tread depth were analyzed: (a) full tread depth, (b) 60% of full tread depth, (c) 30% of full tread depth.
- To evaluate the influence of the tire load on the tire characteristics, five levels of normal load were analyzed: a) 33% of nominal load, (b) 67% of nominal load, (c) 100% of nominal load. (d) 133% of nominal load, (e) 167% of nominal load.

The following sub-sections present results of the sensitivity analysis study.

*Influence of Inflation Pressure on the Tire Characteristics of Interest*

Figure 9 shows the cornering stiffness-inflation pressure dependency curves. The key conclusion reached about the cornering stiffness of a tire is that, an increase in the tire inflation pressure has two counteracting effects:

- A lower cornering stiffness at low vertical loads and a higher cornering stiffness at high vertical loads. These effects are clearly visible in Figure 9. The first effect is caused by the decreasing contact length because of the increased vertical stiffness from the increased inflation. A decrease of contact length (smaller surface area) results in a decrease of cornering stiffness. In the range of high vertical loads, this effect may also be present but it is not dominant.

- A lower inflation pressure and consequently a less stiff carcass results in more rotation of the contact patch, this leads to lower lateral force for the same slip angle, which results in a lower cornering stiffness at high vertical loads. Conversely, a higher inflation pressure and consequently a stiffer carcass results in less rotation of the contact patch. This leads to higher lateral force for the same slip angle, which results in a higher cornering stiffness at high vertical loads.

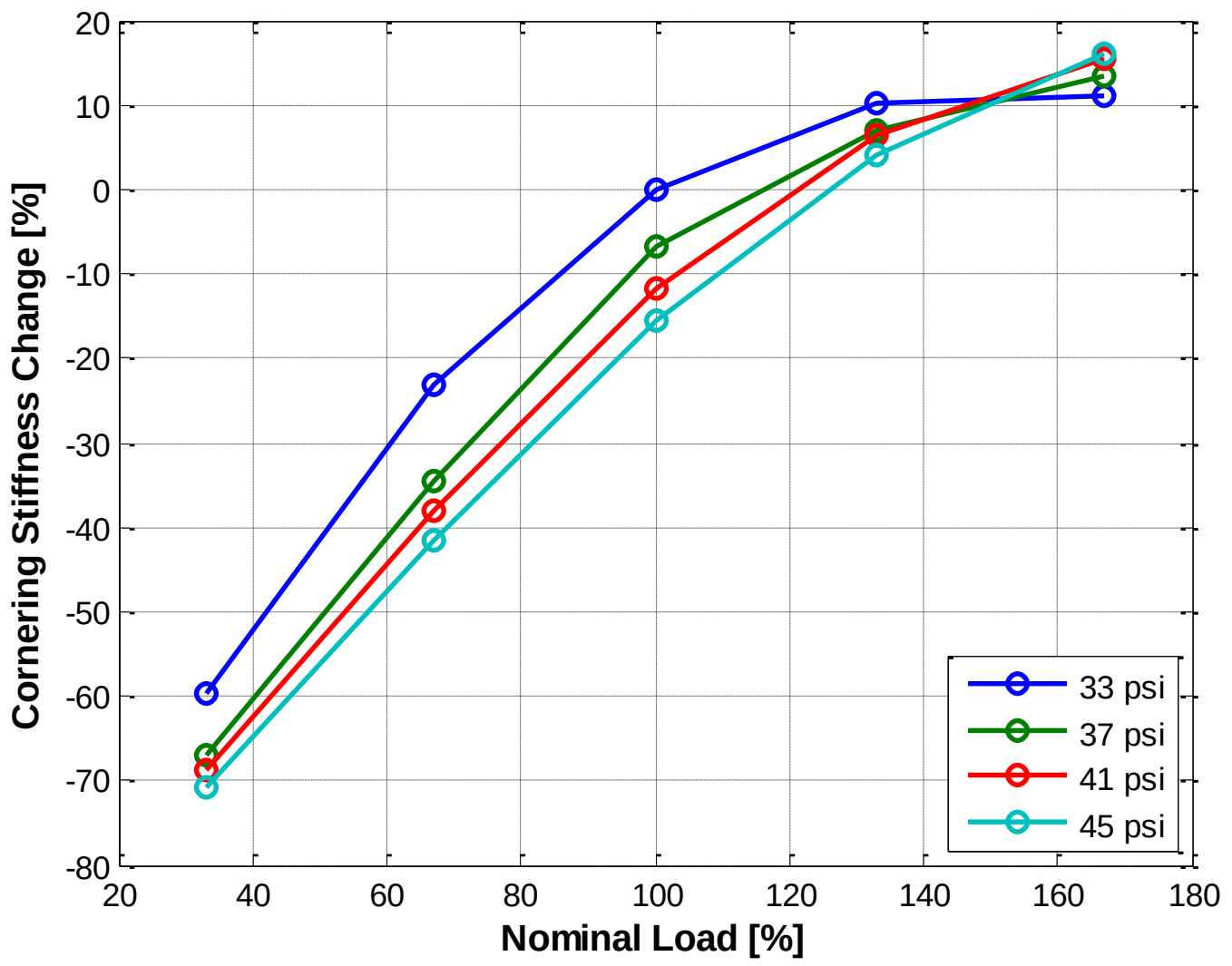

**Figure 9: Influence of inflation pressure on the tire cornering stiffness**

Previous studies [2] have shown that the peak lateral friction characteristics do not show a clear general relation for the inflation pressure. Some tires show a minimum peak lateral friction coefficient at low vertical loads that shifts to an optimum at high vertical loads. Other tires show an opposite behavior. Figure 10 shows the peak lateral friction coefficient-inflation pressure dependency curves. Clearly, the effect of inflation pressure on the peak friction coefficient value doesn't seem to be very dominant.

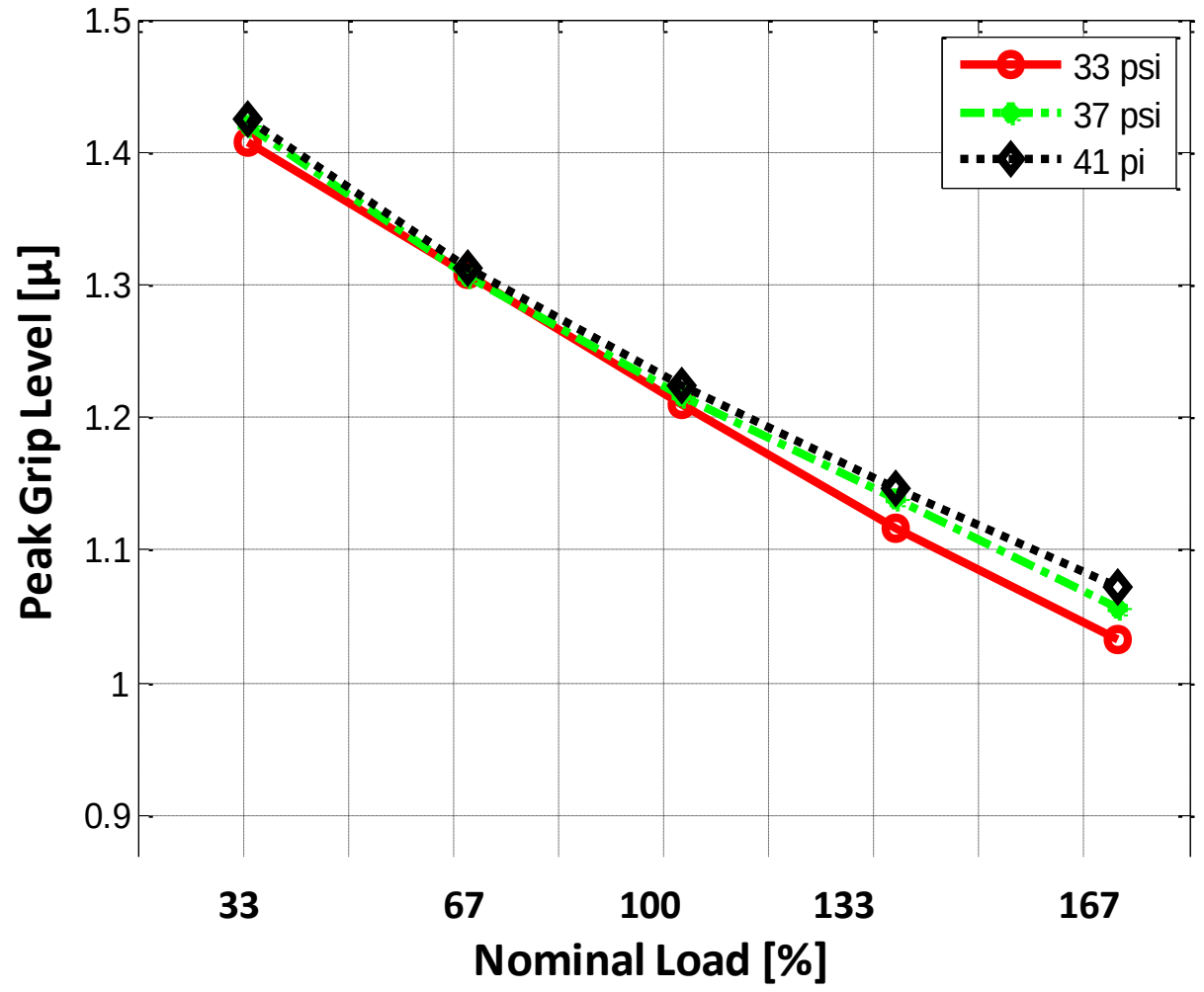

**Figure 10: Influence of inflation pressure on the peak lateral friction coefficient**

## *Influence of Tread Depth (i.e. tire wear state) on the Tire Characteristics of Interest*

As expected, a change in the tire tread depth significantly influences the cornering stiffness characteristics of the tire [Figure 11].

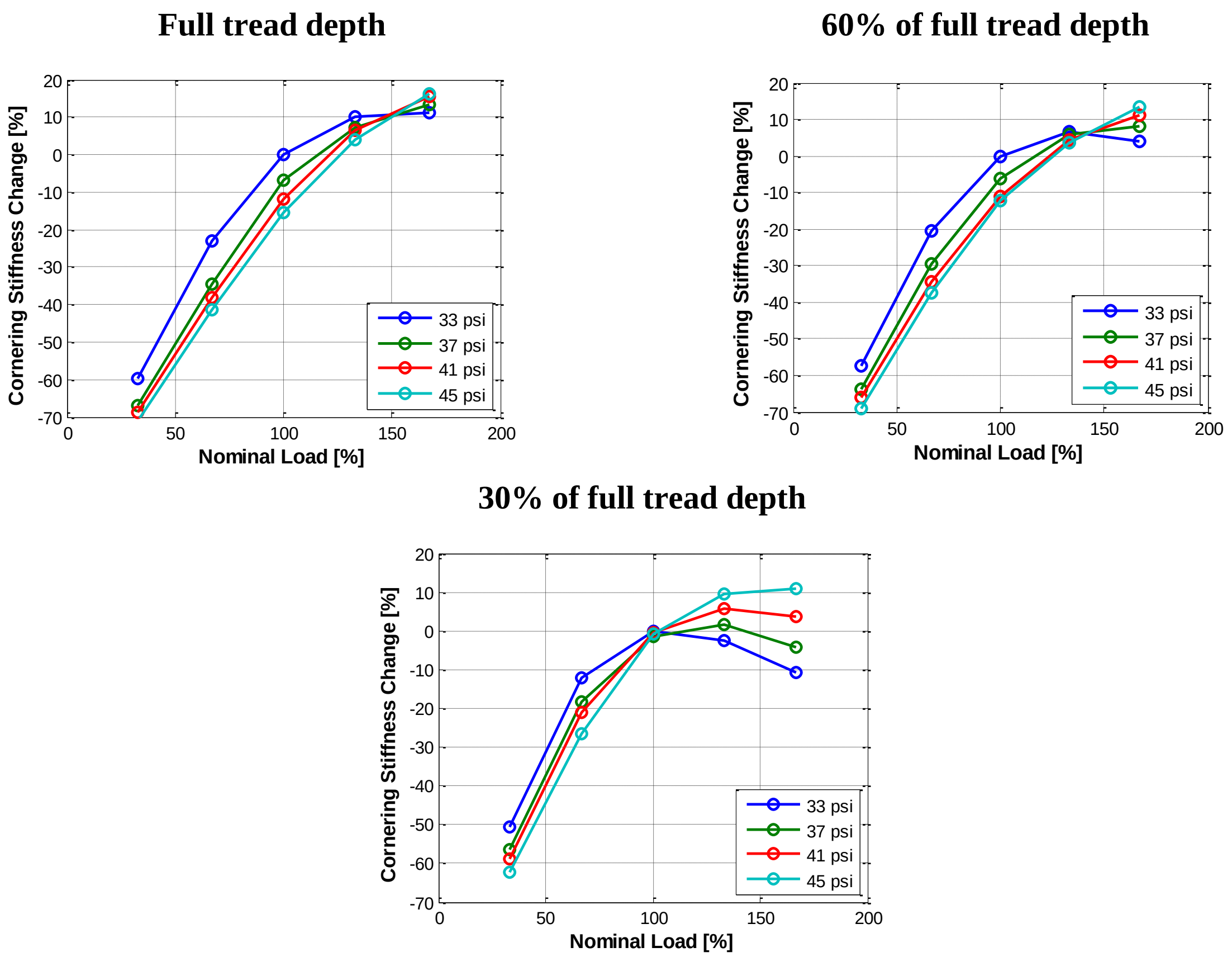

**Figure 11: Cornering stiffness v/s inflation pressure – influence of tire tread depth**

The key conclusions reached with regard to the tread-depth effects are captured in Figure 12 and Figure 13.

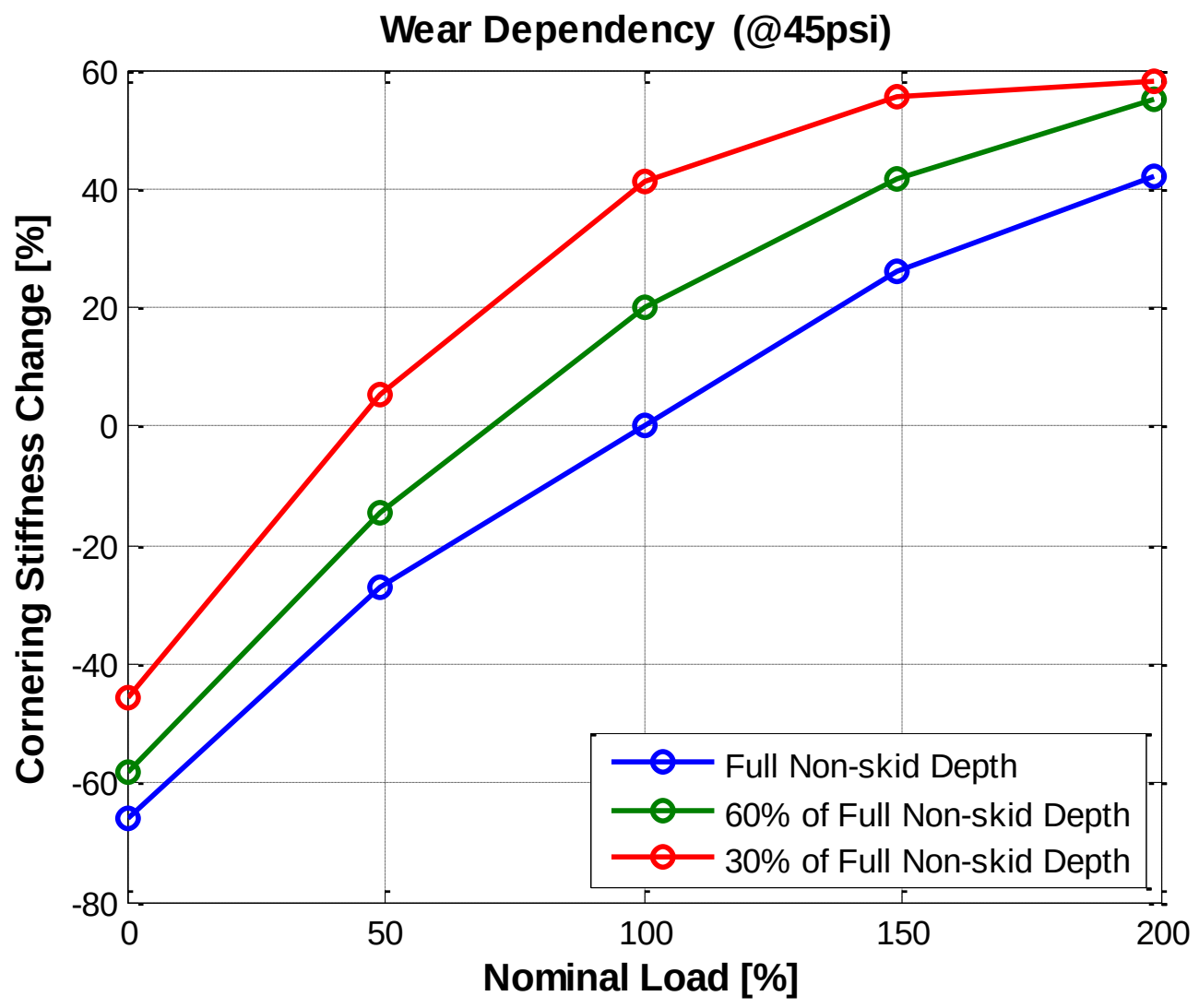

- Lower tread-depth results in a higher cornering stiffness.
- At higher loads, carcass stiffness is the dominant component of cornering stiffness. Hence, even a large change in the tread depth only results in a smaller change in the cornering stiffness.

**Figure 12: Wear dependency effects under constant inflation pressure conditions**

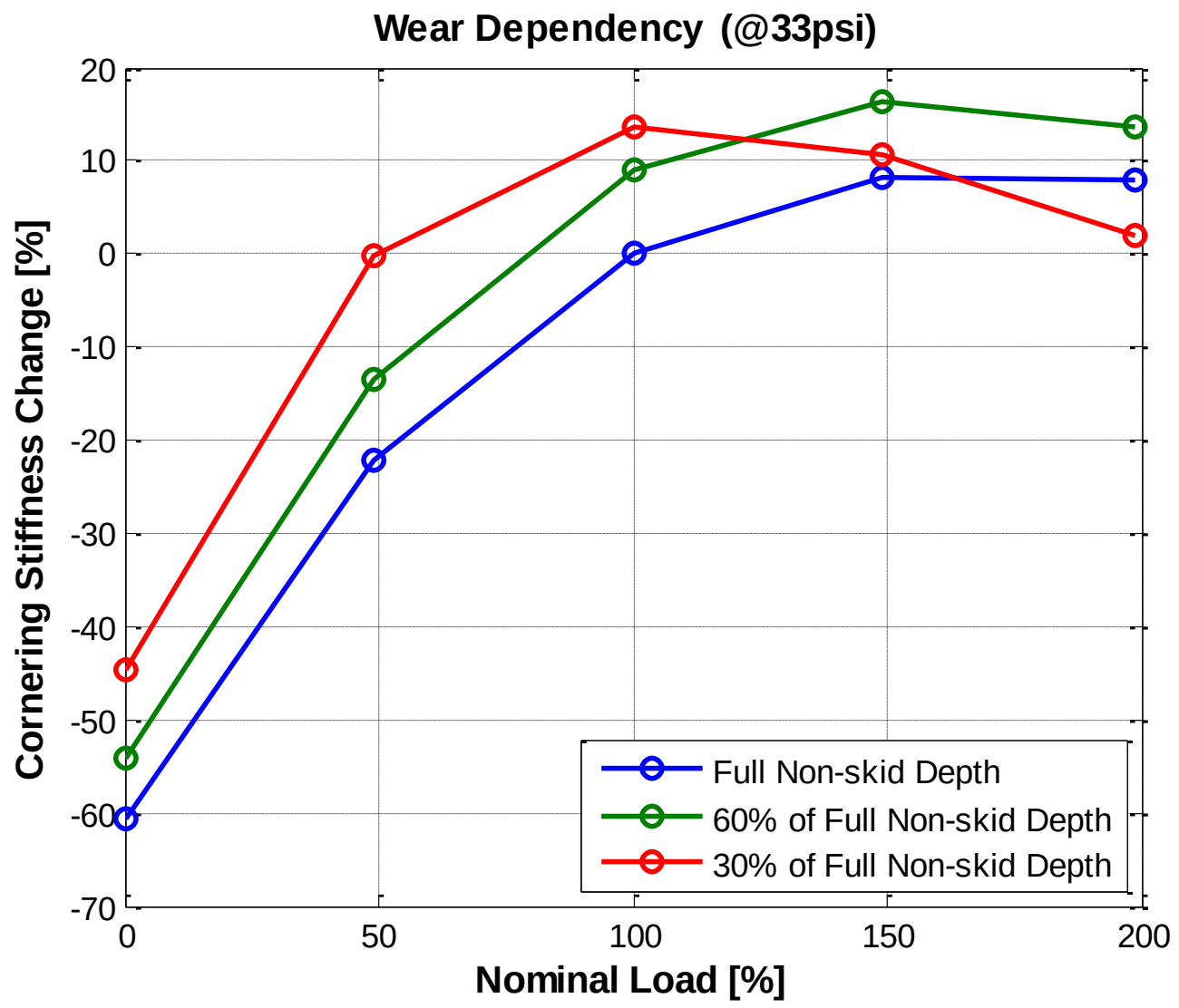

- For a tire with a lower tread depth, the cornering stiffness properties are dominated by the carcass stiffness characteristics.
- As explained previously, lowering the tire inflation pressure decreases the carcass stiffness, which explains the saturation trends seen in the CS curve.
- Furthermore, the saturation starts even earlier for a tire with a lower tread depth due to the dominant effect of carcass stiffness.

**Figure 13: Wear dependency effects under constant inflation pressure conditions**

Tread depth is seen to have a dominant effect on peak friction value [Figure 14].

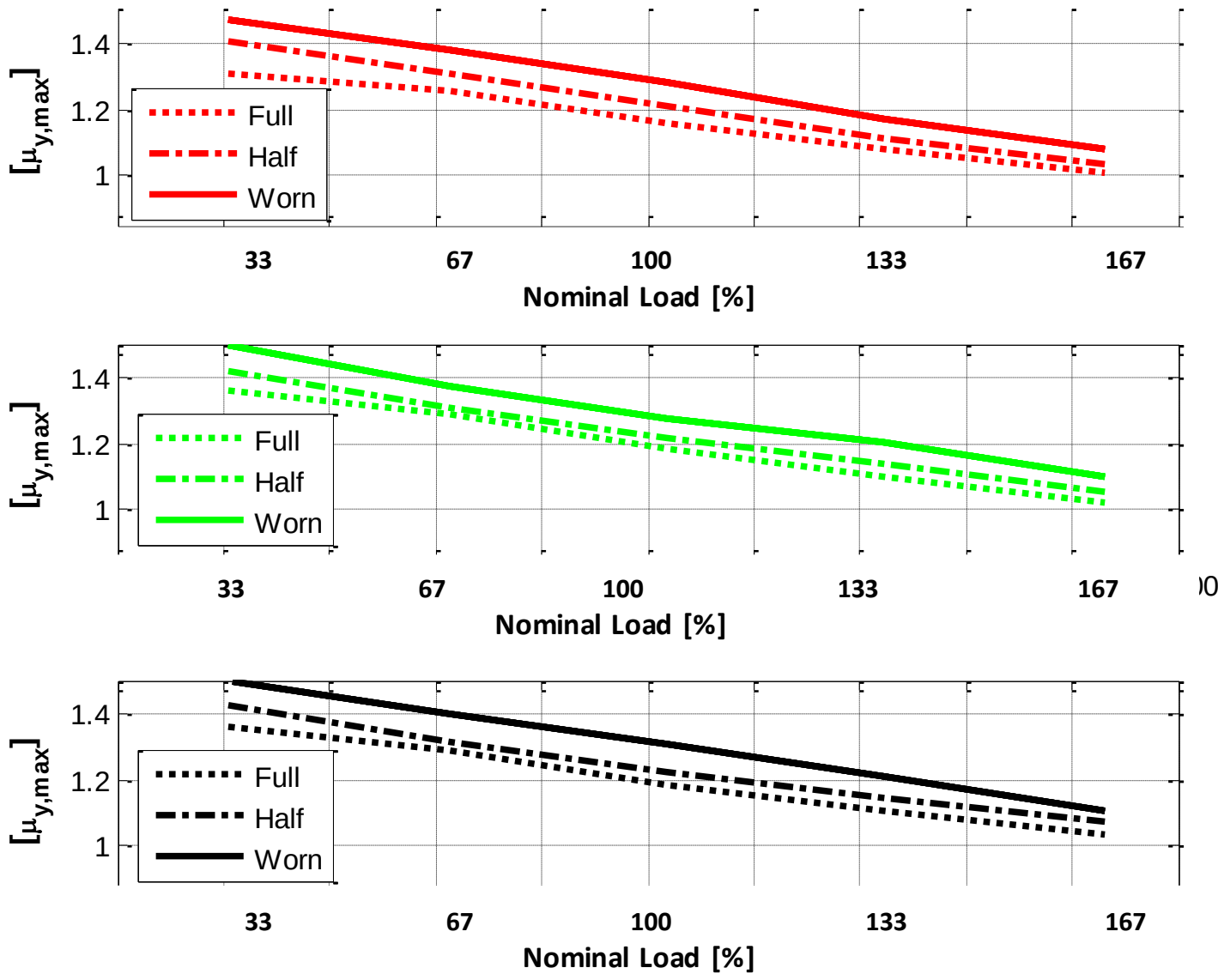

**Figure 14: Peak lateral friction coefficient – tread depth dependency**

This could possibly be attributed to the increase in tread stiffness due to a lower tread depth. Other factors which might influence the peak grip potential of the tire are the net-to-gross ratio of the tire and the localized contact patch pressure, both of which change as a function of the tire wear state and these additional factors can change the trend for a different construction.

### *Influence of Normal load on the Tire Characteristics of Interest*

There are two load dependent parameters affecting the tire cornering stiffness. The first load dependent parameter is the tire contact patch length (L). Typically, L changes with tire load in an order between 1 and 2. The second load dependent parameter is the lateral elastic stiffness, which in turn is due to both the rubber tread elasticity and tire structure lateral compliance.

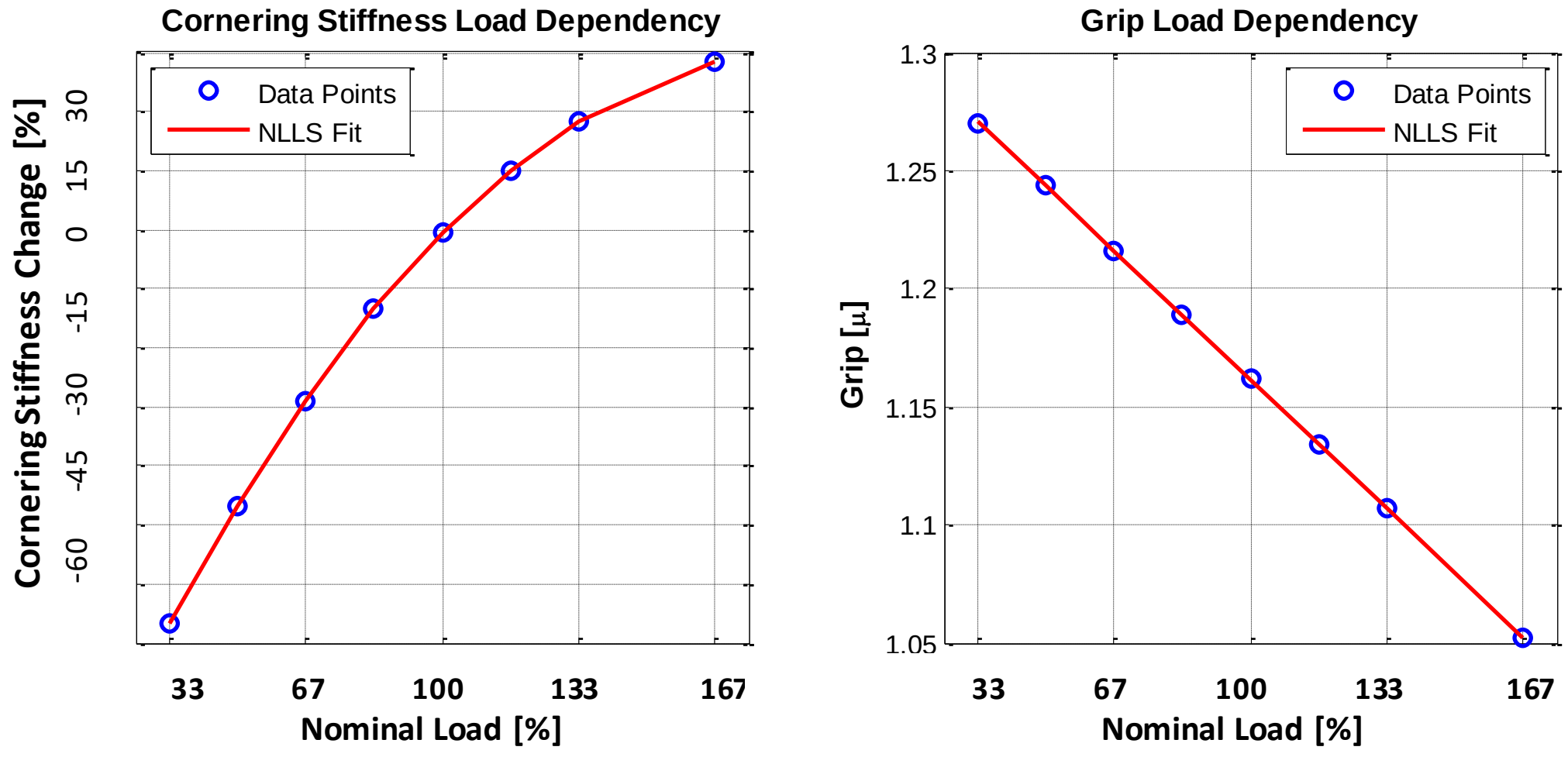

**Figure 15: Influence of normal load on: (a) the tire cornering stiffness, (b) peak grip level**

A final observation stemming from the results in Figure 15 is that the friction coefficient decreases with increase of tire load. The load dependence of the friction coefficient includes the effect of the rubber modulus. A secondary source of load dependence can be the friction induced thermal heating in the contact area leading to a reduction of the friction coefficient with increasing load [Figure 16].

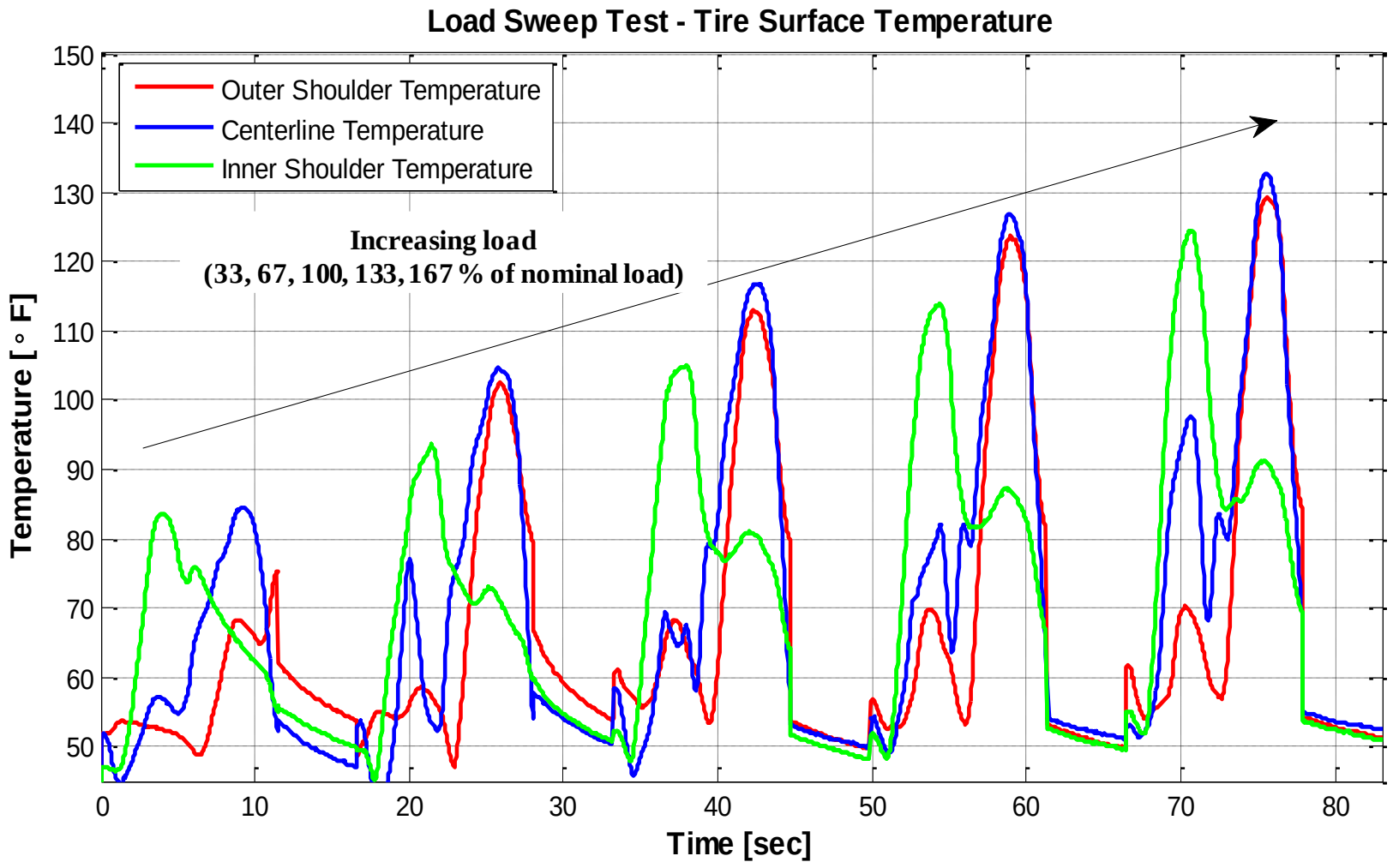

**Figure 16: Load sensitivity – tire surface temperature**

This load dependence is not due to an intrinsic pressure dependence of the rubber friction coefficient, but a kinetic effect related to the build-up of the flash temperature [11] in rubber–road asperity contact regions during slip.

### *Influence of Temperature on the Tire Characteristics of Interest*

A second set of tests were run on the Flat-Trac® machine to include the influence of temperature. The procedure consisted of running the test on new tires without a warm-up or break-in procedure. The tires were tested through 23 slip sweeps at 1 load and 1 camber condition. IR pyrometers were installed on Flat-Trac® machine allowing tire temperatures to be recorded throughout each test [Figure 17].

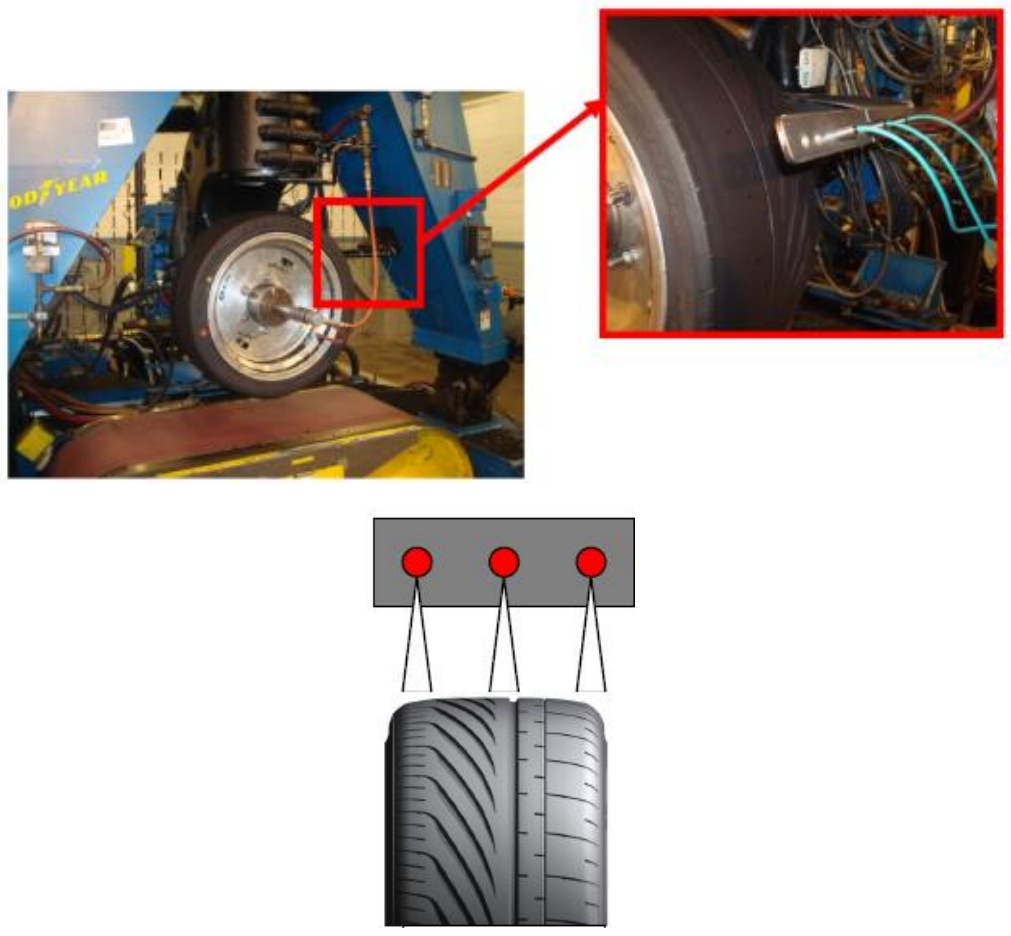

**Figure 17: Tire temperature apparatus installed on the Flat-Trac® test machine**

The IR pyrometers give the machine the capability to record tire shoulder and centerline surface temperatures. Figure 18 shows the cornering stiffness and peak grip-temperature dependency curves.

(a)

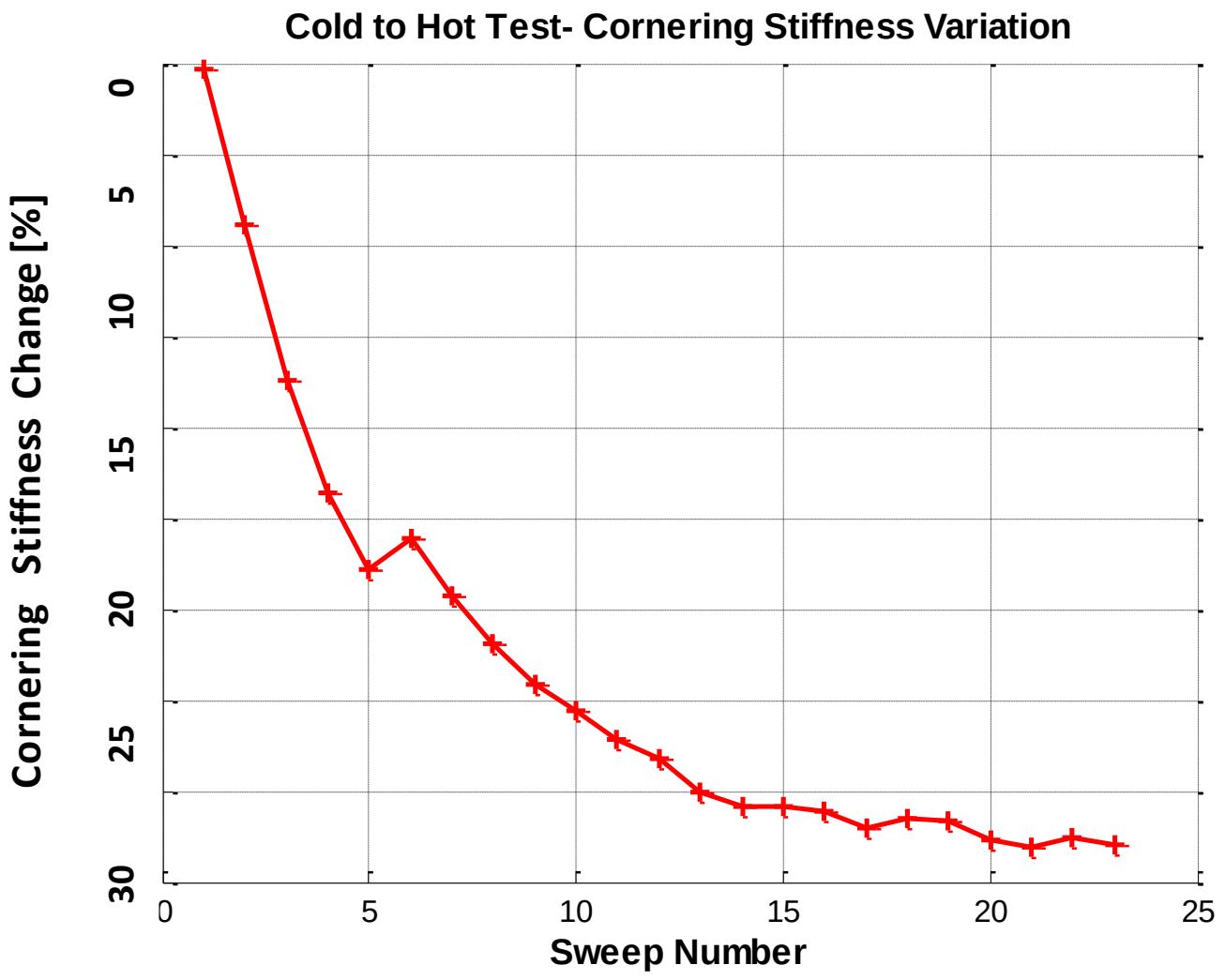

(b)

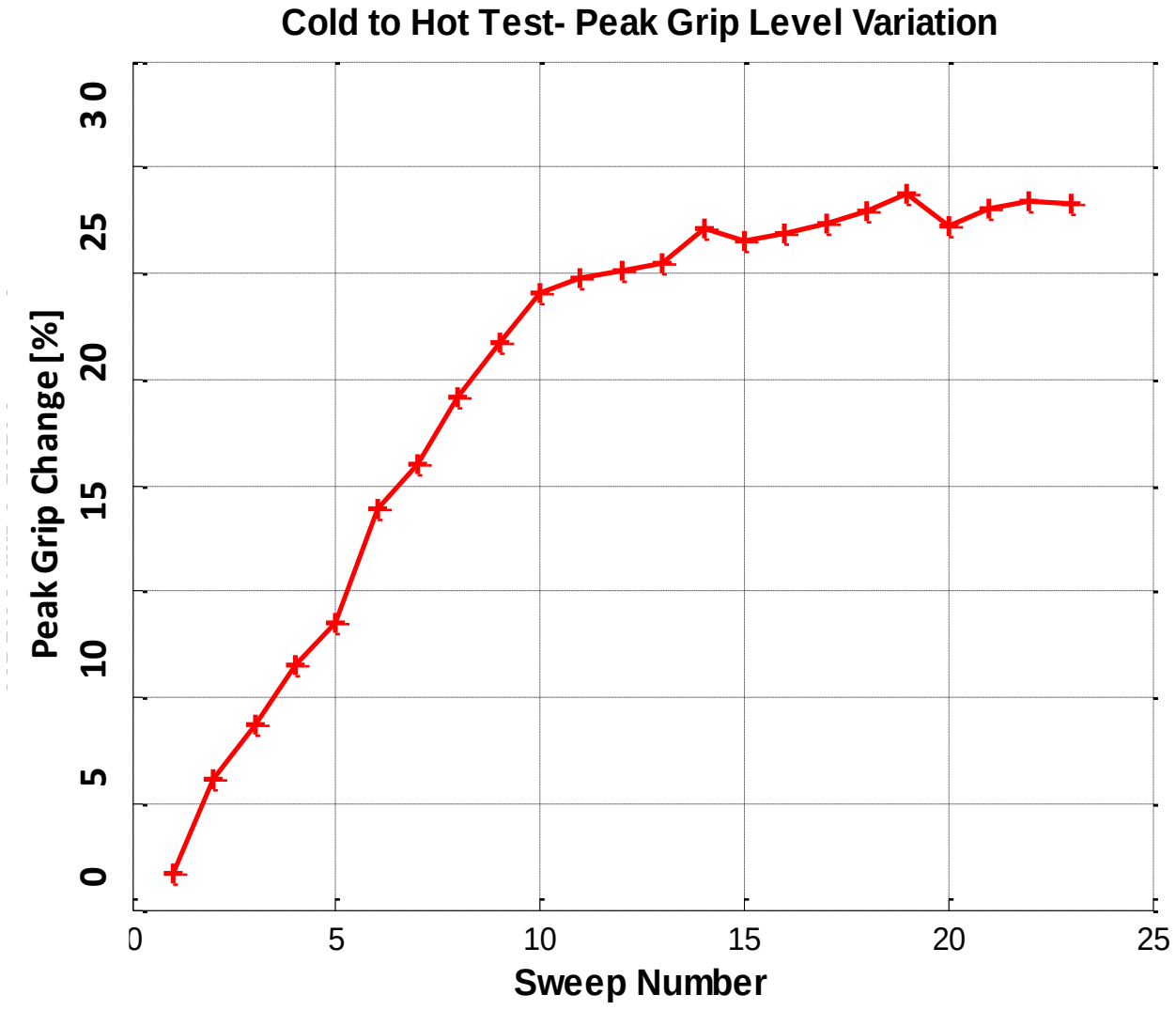

**Figure 18: Influence of temperature on: (a) the tire cornering stiffness, (b) peak grip level**

The influence of temperature can be mainly attributed to two main visco-elastic properties of rubber, which change with temperature as shown in
Figure 19:

1. The storage modulus or tread rigidity, which influences cornering stiffness. This changes due to the bulk temperature of the tire.
2. The coefficient of friction decides the peak lateral grip of the tire. This parameter is only influenced by the surface temperature of the tire at the road interface. Typically, the interaction forces reach their maximum values only within a narrow temperature range, while decay significantly outside of it [Figure 19].

(a)

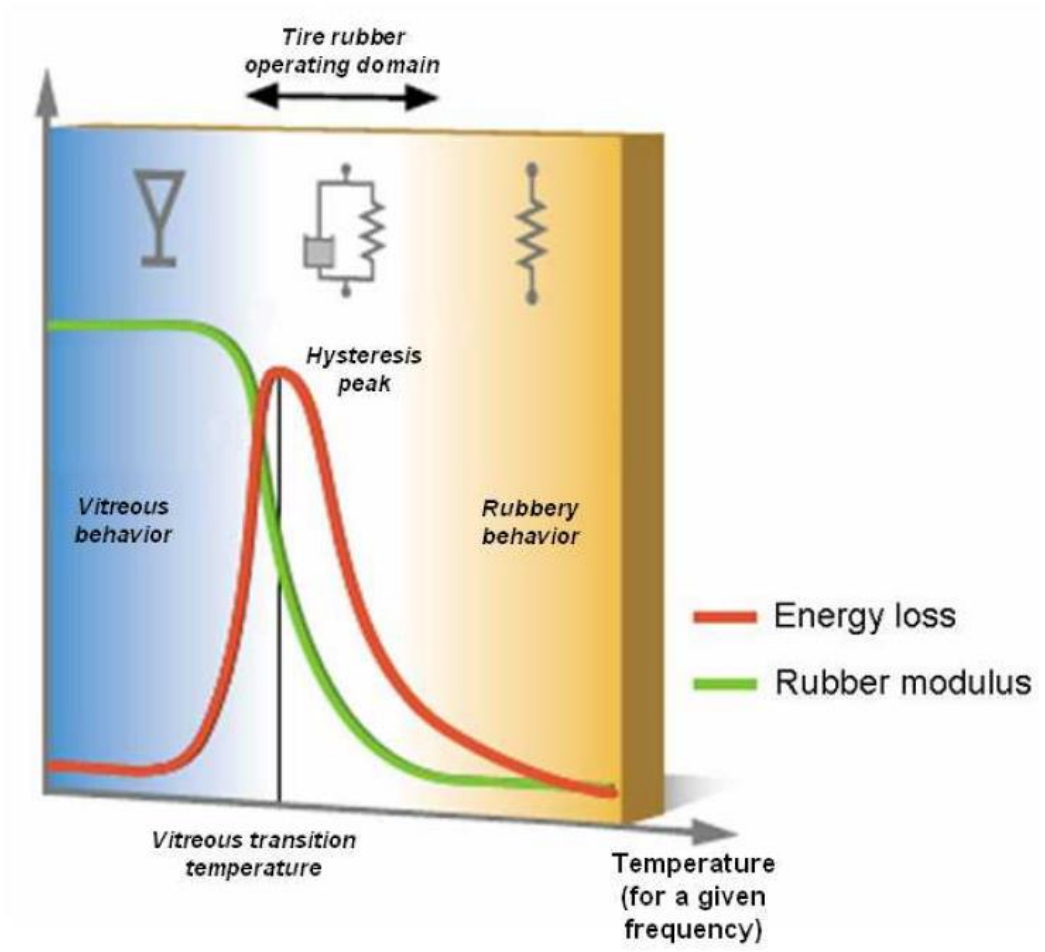

(b)

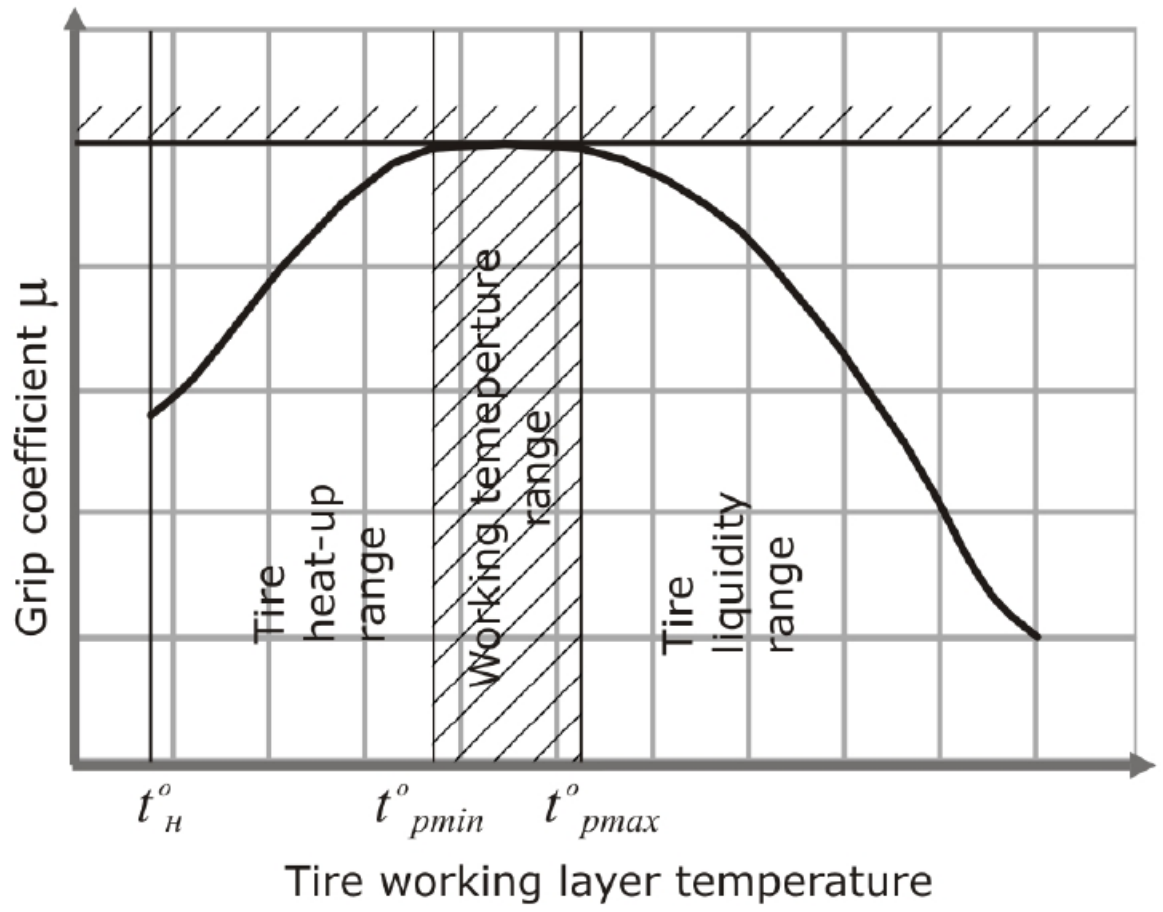

**Figure 19: Temperature influence on: (a) rubber properties [5] and, (b) the resultant influence on friction coefficient [9]**

Table 2 presents a summary of the sensitivity analysis, determining the influence of variations in the tire inflation pressure, tread depth, normal load, and temperature on the tire cornering stiffness and the peak friction force coefficient.

**Table 2: Summary of dependencies**

| | Tire Surface Temperature | Tire Bulk Temperature | Inflation Pressure | Normal Load | Rolling Speed | Tread Depth |
|---|---|---|---|---|---|---|
| **Cornering Stiffness (CS)** | High Dependency | High Dependency | High Dependency | High Dependency | Negligible Dependency | High Dependency |
| **Peak Grip Level (μ)** | High Dependency | Low Dependency | Low Dependency | High Dependency | High Dependency | High Dependency |

A quantifiable measure of the influence of the tire operating conditions on the cornering stiffness and the peak grip friction level is summarized in Table 3.

**Table 3: Sensitivity analysis- Summary**

| **Tire Type** | | **Factors Influencing Tire Characteristics** | | |
|---|---|---|---|---|
| | | Pressure | Tread Depth | Temperature |
| **Summer Tire (High Performance)** | Cornering Stiffness | **10% increase with a 20% change in inflation pressure from nominal conditions** | **30% increase with a 60% decrease in tread depth** | **20-25% drop from cold to hot tire conditions (* strongly influenced by the tire bulk temperature)** |
| | Peak Grip | **Influence doesn't seem to be very dominant** | **10% increase with a 60% decrease in tread depth** | **10% drop from cold to hot tire conditions** |

The next section of this paper presents details about an improved magic formula (MF) tire model adapted to cope with changes in the tire operating conditions.

## 4. Magic Formula (MF) Adaptation

This section describes extensions to the widely used magic formula tire model. More specifically, extensions have been made to the magic formula expressions for tire cornering stiffness and peak grip level, details of which are described in the subsequent sub-sections.

### Cornering Stiffness Adaptation

Based on the Pacejka tire model formulation, the expression for tire cornering stiffness (BCD) is given as:

$$BCD = a_3 \cdot sin\ (2\, arctan\left(\frac{F_z}{a_4}\right)) \qquad (8)$$

where

$a_3$ = maximum cornering stiffness (at camber angle $(\gamma) = 0$)

$a_4$ = load at maximum cornering stiffness

In this study, the parameters 'a3' and 'a4' were calculated using the nonlinear least-squares (NLLS) curve fitting algorithm. The curve fitting results are shown in Figure 20.

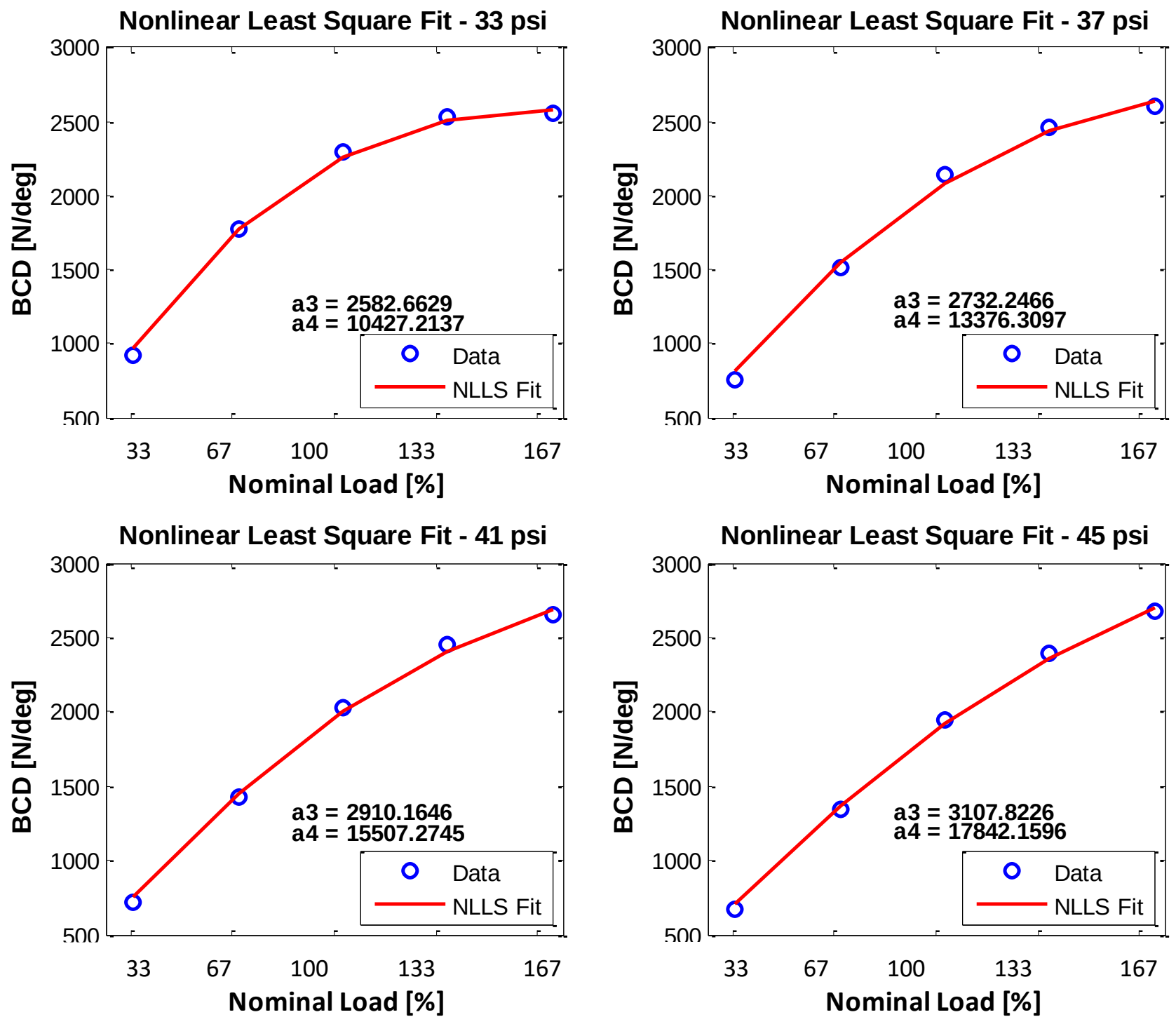

**Figure 20: NLLS Fitting Results - ‘a3’ and ‘a4’ coefficient**

To capture the pressure dependency, a second-order pressure scaling term was used for the parameter ‘a3’, and a first-order pressure scaling term for the parameter ‘a4’, as shown in expression below.

$$BCD = (\, a_{31} * x^2 + \; a_{32} * x + a_{33})\,.\,sin\;(2\,arctan\left(\frac{F_z}{(\,a_{41} * x + \; a_{42})}\right) \qquad (9)$$

*where x=pressure*

Model fitting results are shown in Figure 21.

**‘a3’ pressure adaptation**

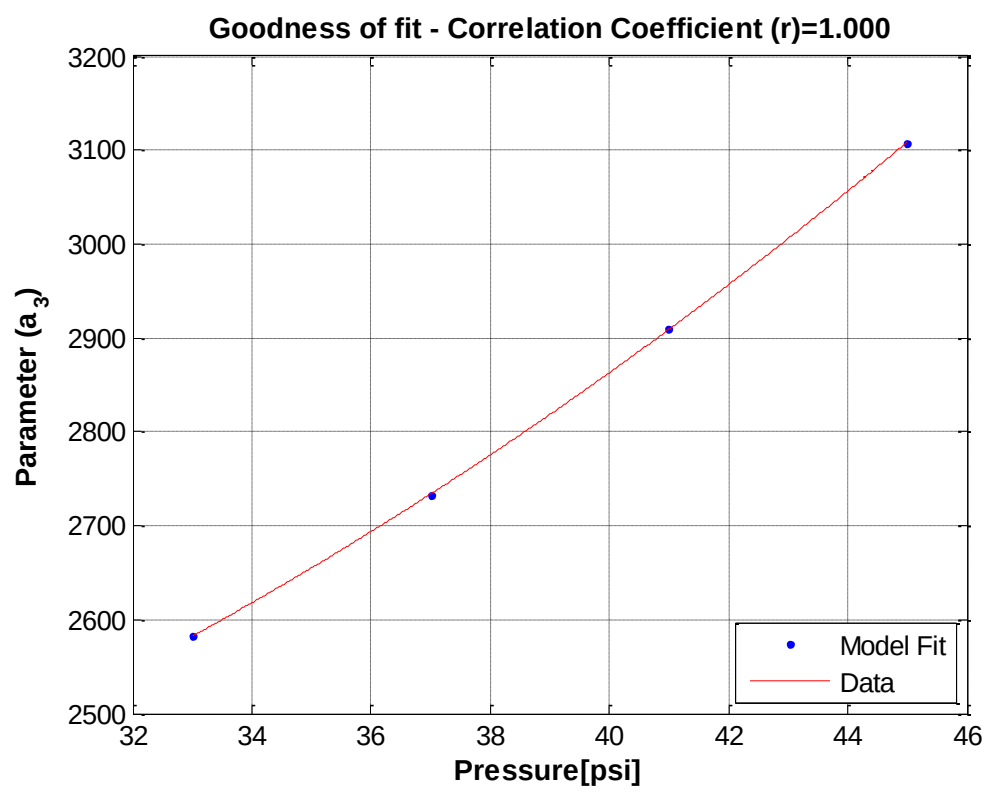

**‘a4’ pressure adaptation**

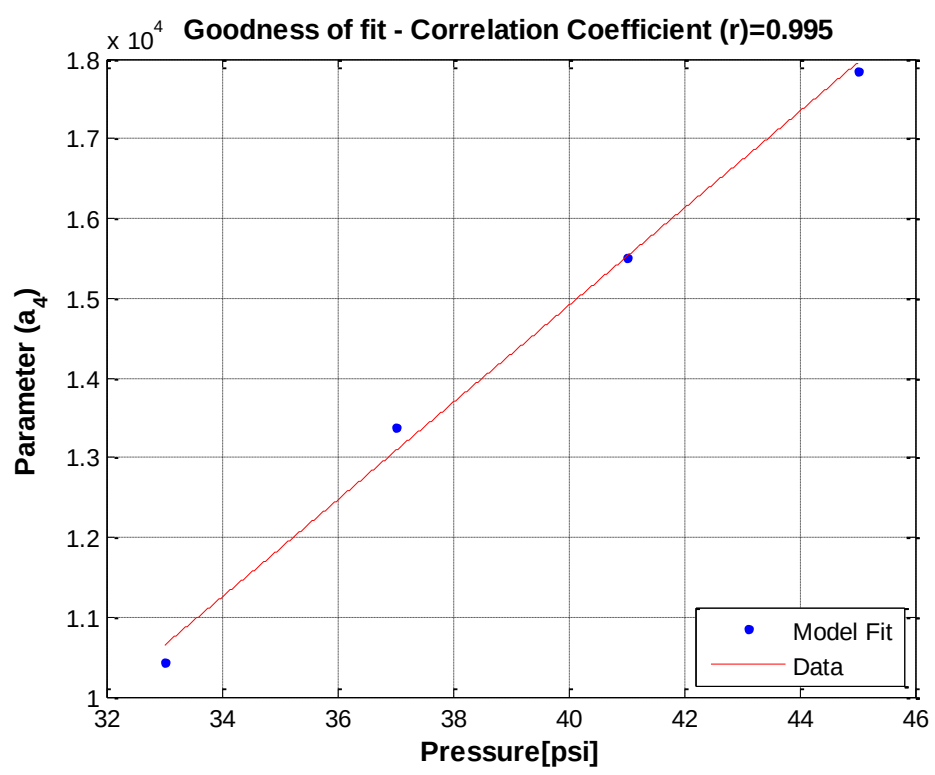

**Figure 21: Model Fitting Results**

The above analysis was repeated for tires with different levels of tread depth, with the underlying objective of capturing both the inflation pressure and tread depth effects simultaneously. To capture the tread depth dependency, second-order scaling terms were used to adapt the parameters ‘a31’, ‘a32’, ‘a33’, ‘a41’ and ‘a42’ to tread depth changes. Shown below in Equation (10) is the modified expression for cornering stiffness with inflation pressure and tread depth adaptation terms.

$$BCD = ((a_{311} * y^2 + a_{312} * y + a_{313}) * x^2 + (a_{321} * y^2 + a_{322} * y + a_{323}) * x + (a_{331} * y^2 + a_{332} * y + a_{333})) \cdot \sin\left(2\,arctan\left(\frac{F_z}{((a_{411} * y^2 + a_{412} * y + a_{413}) * x + (a_{421} * y^2 + a_{422} * y + a_{423}))}\right)\right) \qquad (10)$$

*where x=pressure & y=tread depth*

The performance of adapted model was tested against measurement data and it showed good performance for the full range of inflation pressures and tread-depths. As an illustrative example, model fitting results for a high-performance summer tire are shown in Figure 22.

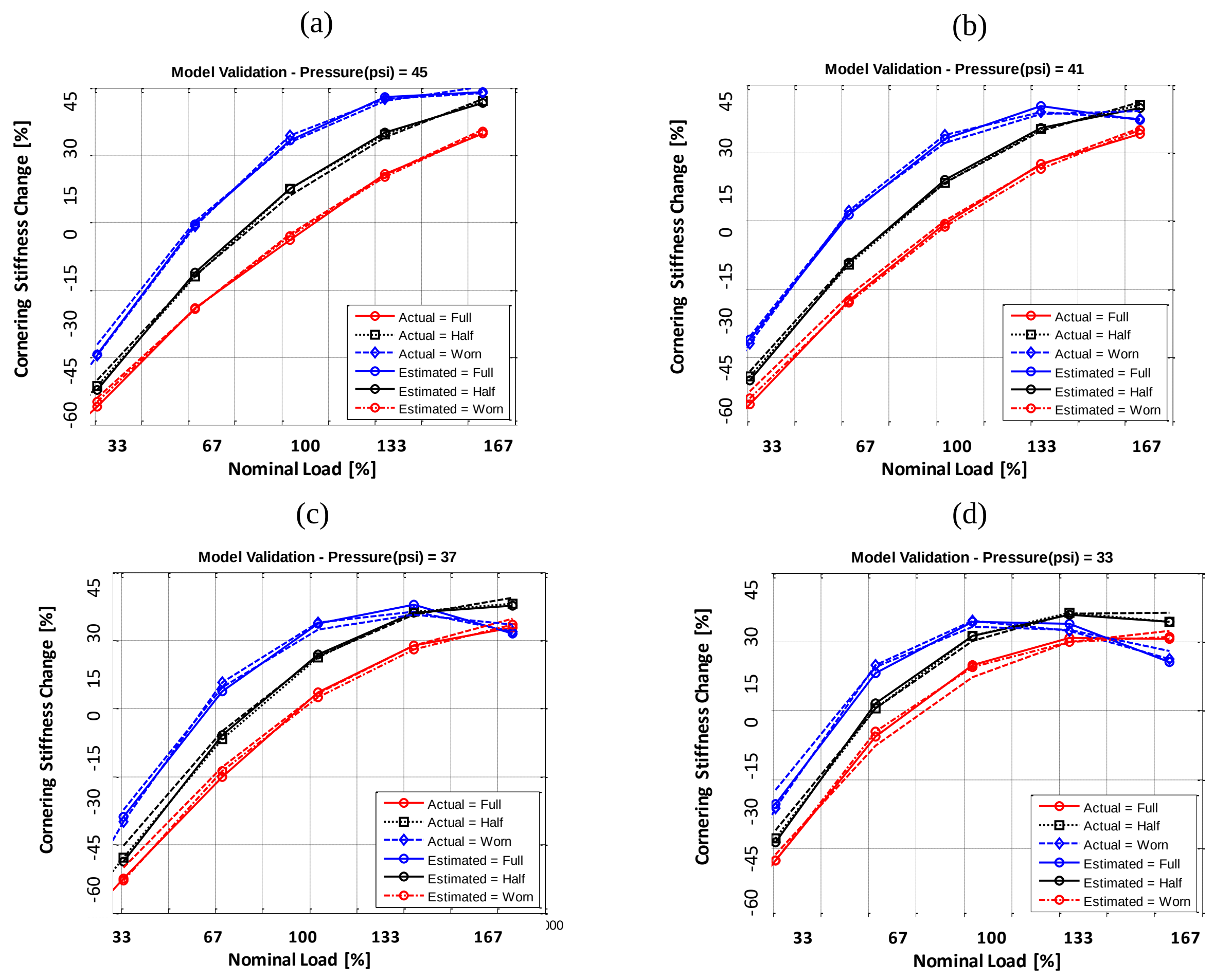

**Figure 22: Tire cornering stiffness: actual (measurement) v/s estimated (using adapted model) Model fitting performance at: (a) 45 psi, (b) 41 psi, (c) 37 psi, (d) 33 psi**

As a final step, a second-order temperature dependent scaling factor was used to cope with the changes in the tire temperature [Figure 23]. More specifically, the term “tire temperature" as used herein refers to the tire surface temperature.

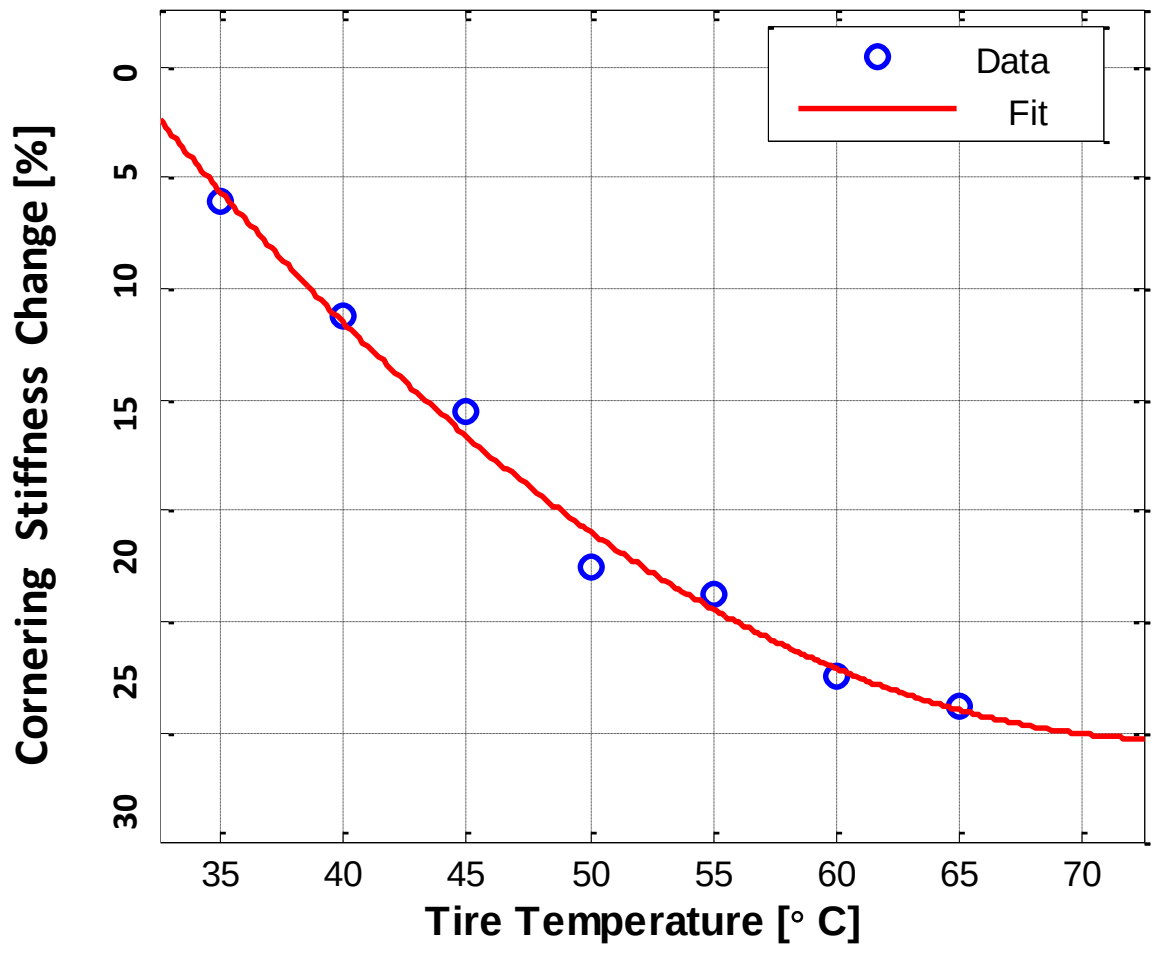

**Figure 23: Model fitting results- temperature dependency**

The final expression for the adapted cornering stiffness term for the Pacejka tire model formulation is shown below.

$$BCD = ((a_{311} * y^2 + a_{312} * y + a_{313}) * x^2 + (a_{321} * y^2 + a_{322} * y + a_{323}) * x + (a_{331} * y^2 + a_{332} * y + a_{333})) \cdot \sin\left(2\, arctan\left(\frac{F_z}{((a_{411} * y^2 + a_{412} * y + a_{413}) * x + (a_{421} * y^2 + a_{422} * y + a_{423}))}\right)\right) \cdot (b_{11} * z^2 + b_{12} * z + b_{13}) \quad (11)$$

*where x=pressure, y=tread depth & z=temperature*

The model inputs include: tire inflation pressure, tread depth, tire load, tire temperature, and tire ID (required for using tire specific model coefficients). It is envisioned by tire researchers [20-24] that in the coming years, the intelligent tire concept using tire attached sensor modules will have the capability provide information about the tire inflation pressure, load, tread depth and tire ID information (required for using tire specific model coefficients). Temperature information available from tire mounted sensor systems typically consists of tire cavity air pressure temperature and/or tire inner-liner temperature. The temperature measurement of interest here for model adaptation purposes is the tire surface temperature, i.e. the temperature at the tire-road interface. It is proposed to use an empirical model to predict the tire surface temperature. The model requires the following inputs:

- Inner liner temperature (available from tire attached sensor systems)
- Ambient Temperature (from vehicle CAN)
- Frictional Energy (estimated using vehicle CAN signals)
- Forward Velocity (from vehicle CAN)

- Temperature at previous time-step (internal model calculation)

An artificial neural network based model was used to fit the empirical model for tire surface prediction. As an illustrative example, a 2-layer recurrent neural network model with 14 neurons is shown here. This model has been trained per-axle using experimental data from hot laps to estimate tire surface temperature. In this, the carcass (tire inner-liner) temperature was collected from wheel-mounted IR sensors for experimental purposes. This temperature is expected to be provided by tire-based sensors in the future. Other parameters such as ambient temperature, vertical load, forward velocity, slip angles are assumed to be available from vehicle-based sensors and estimators. Frictional energy input is based on lateral, longitudinal sliding forces and slip velocities as shown below.

$$E_x = F_{sx} V_{sx}$$
$$E_y = F_{sy} V_{sy} \tag{12}$$

The slip velocities are calculated from approximate values of slip angles and slip ratios. The lateral and longitudinal forces are approximated based on accelerations from vehicle IMU and vertical load calculated based on load transfer as shown below.

$$\begin{aligned} V_{sx} &= V_x - r\omega \\ V_{sy} &= V_y = V_x \tan\alpha \\ F_x &= F_z \frac{a_x}{g} \\ F_y &= F_z \frac{a_y}{g} \end{aligned} \tag{13}$$

The proposed structure of the model is shown in Figure 24.

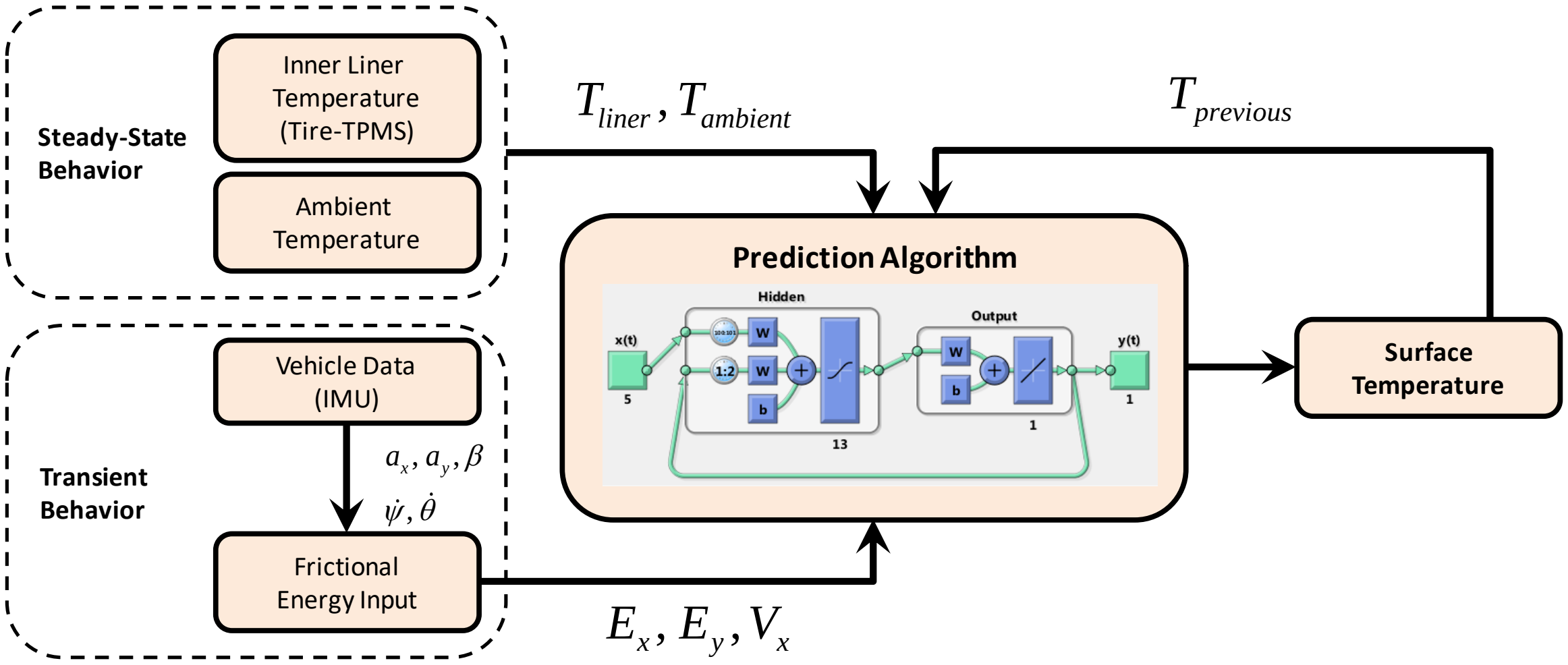

**Figure 24: Surface temperature prediction model - flowchart**

For the purpose of experimental validation of the model, vehicle data from a maximum speed lap performed on another day was utilized and the results of the prediction model are shown in Figure 25 for the front-right tire and the rear-left tire respectively. Although this approach can yield satisfactory prediction of surface temperature as shown in Figure 25, the accuracy of the model is highly dependent on the availability of reliable training data. The measurement of carcass temperature by tire attached sensors also plays a crucial role as any errors can accumulate errors in the system. A more robust approach would involve utilizing a semi-empirical model that can be fitted to experimental data.

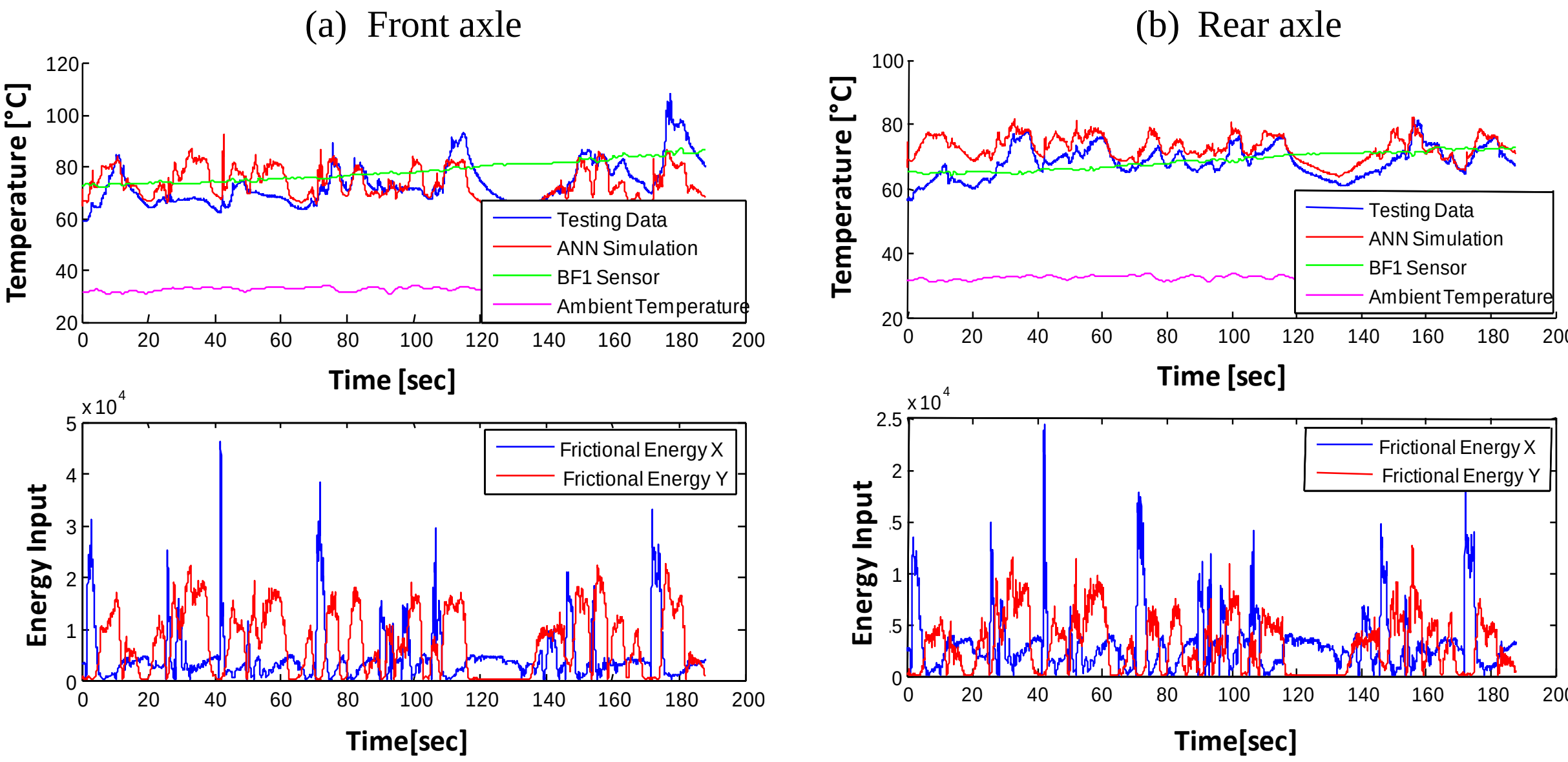

**Figure 25: Tire surface temperature prediction results: (a) front axle, (b) rear axle**

### Peak Grip Adaptation

Based on the Pacejka tire model formulation, the expression for the peak lateral force (D) is given as:

$$D = a_1 \cdot F_z + a_2 \tag{14}$$

where
$a_1$ = Load dependency on lateral friction
$a_2$ = Lateral friction level

Shown below in Equation (15) is the modified expression for the peak lateral friction coefficient with tread depth adaptation.

$$\mu_{y_{max}} = (a_{11} * y^2 + a_{12} * y + a_{13}) * Fz + (a_{21} * y + a_{22}) \tag{15}$$

*where y= tread depth*

A second-order temperature scaling factor was used to cope with the changes in the tire temperature [Figure 26].

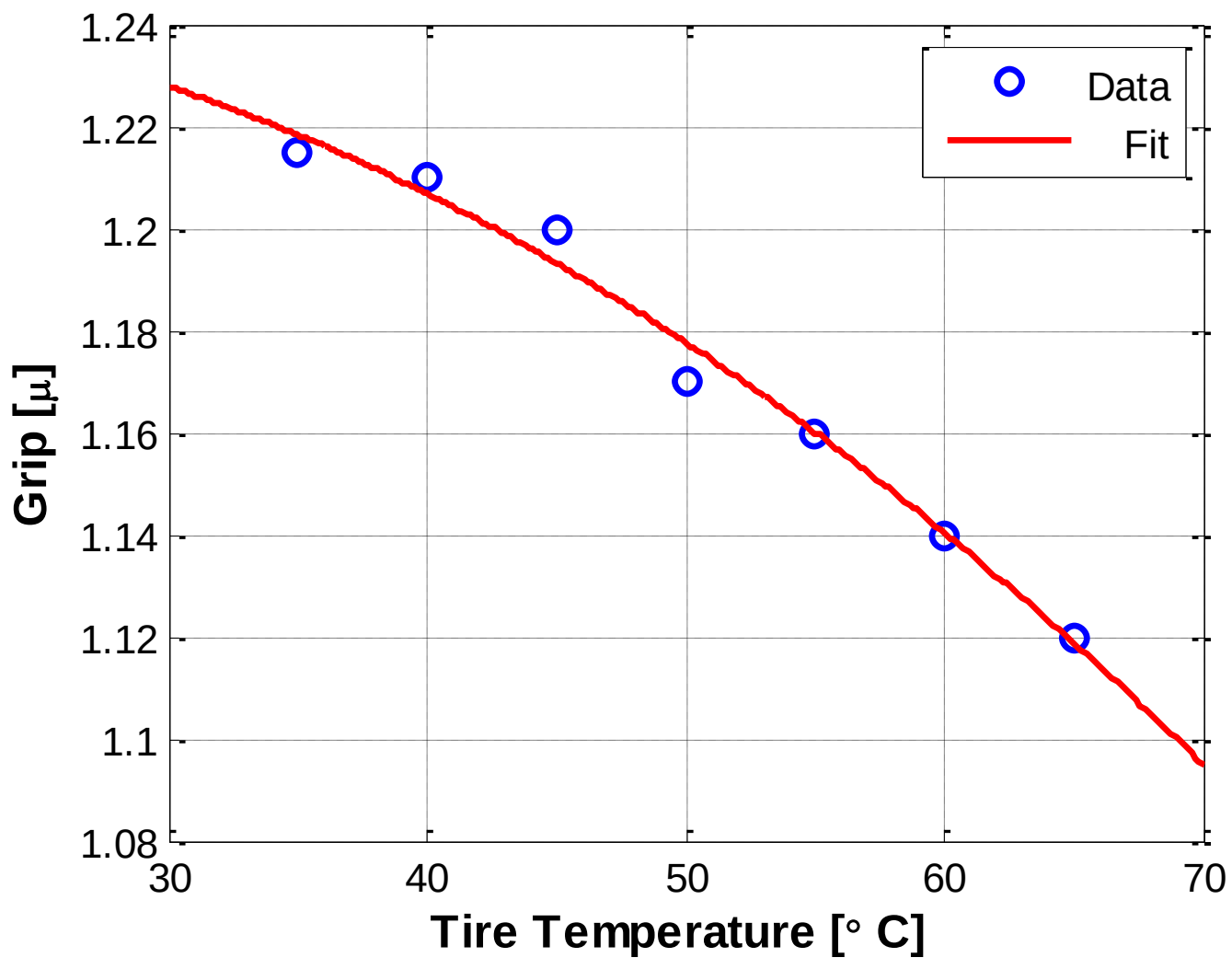

**Figure 26: Model fitting results- temperature dependency**

The final expression for the adapted peak grip level term for the Pacejka tire model formulation is shown below.

$$\mu_{y_{max}} = (a_{11} * y^2 + a_{12} * y + a_{13}) * Fz + (a_{21} * y + a_{22}) * (b_{11} * z^2 + b_{12} * z + b_{13}) \qquad (16)$$

*where y= tread depth & z= temperature*

Since the tire model used in this study was the Pacejka tire model, the result is a set of adaptation equations for the Pacejka model given by Equations (11) & (16), introduced to consider the influence of inflation pressure, tread depth and temperature on the tire force curve [Figure 27].

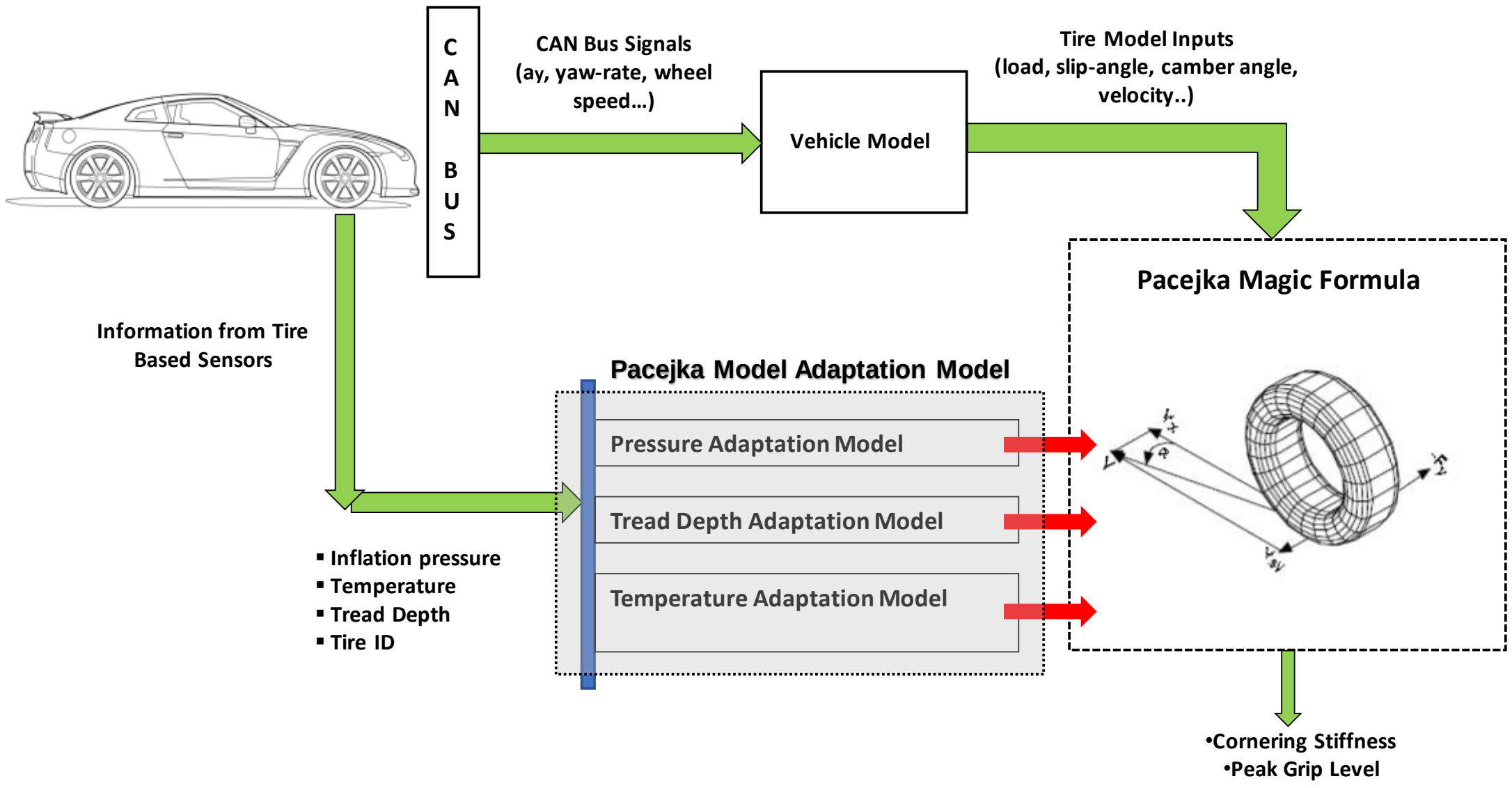

**Figure 27: Model implementation flowchart**

A possible implementation of such an adaptation model in the vehicle ECU is illustrated in Figure 28.

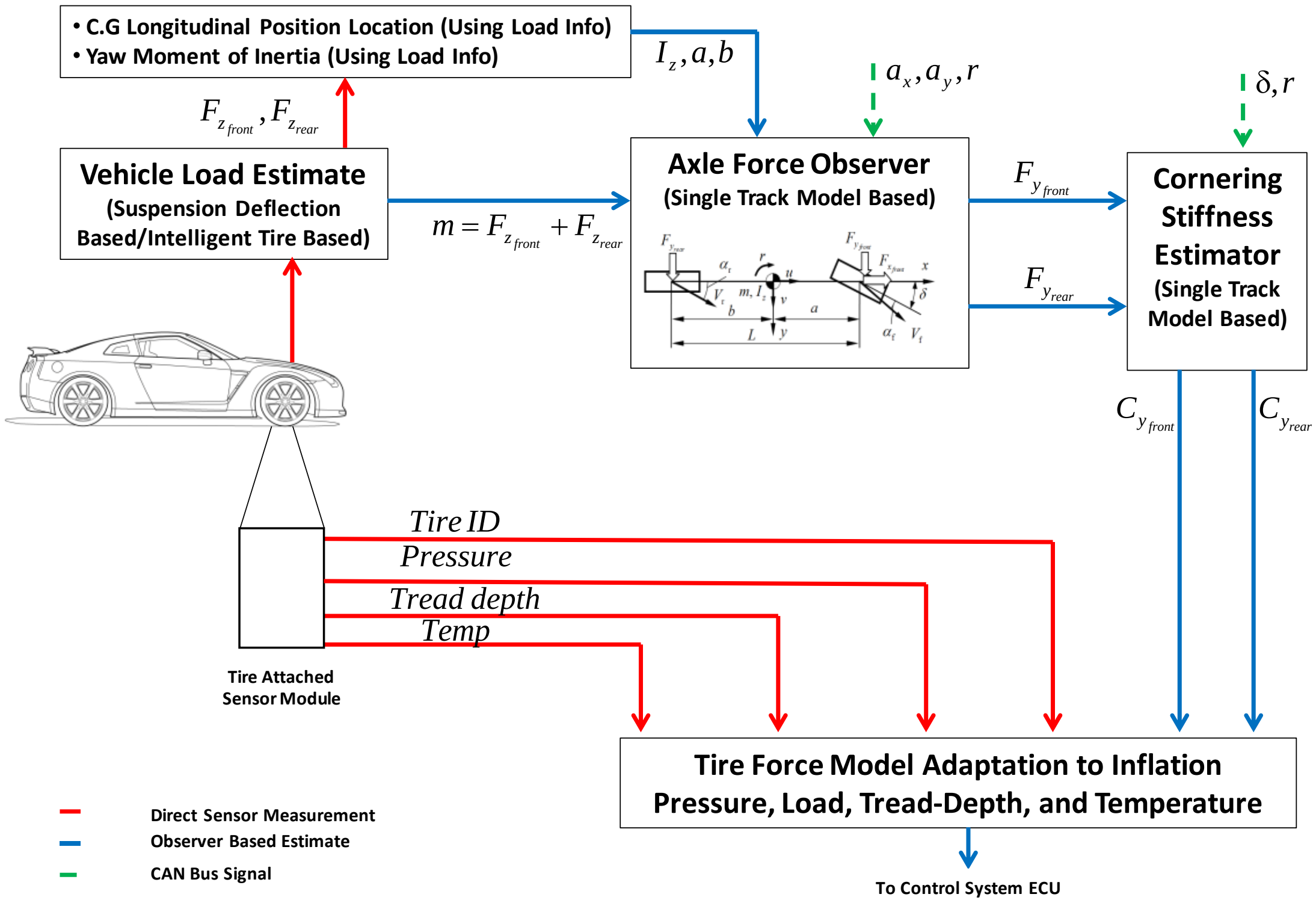

**Figure 28: Tire force model adaptation with tire sensed information**

It is noteworthy to mention that one needs to also consider compliance effects due to the suspension and steering system and the road roughness effects [Figure 29].

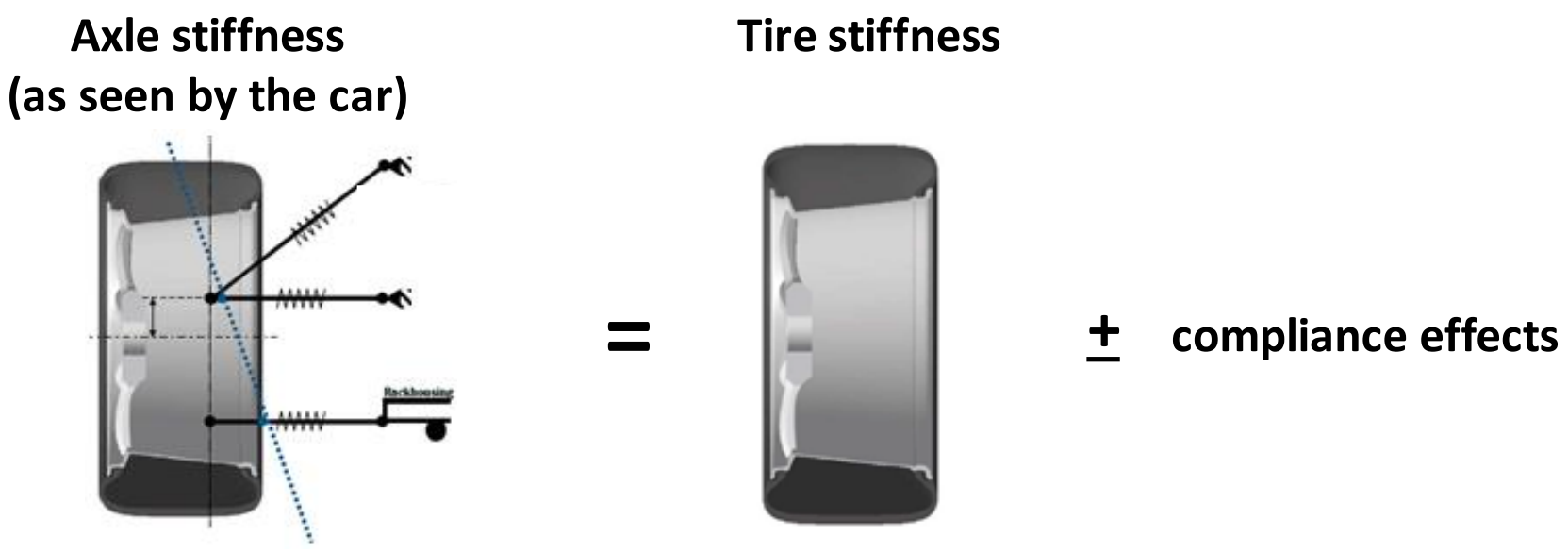

**Figure 29: Axle stiffness, as seen by the vehicle**

This is so because the stiffness seen by the vehicle axle comprises of the tire cornering stiffness and the compliance effects. Hence the authors propose using a cornering stiffness estimator [38] based on a bicycle model [Figure 30]. The on-vehicle estimator uses signals available on the vehicle CAN network to provide a low frequency long term update of the cornering stiffness. Furthermore, information from the tire attached sensor module is used to compensate this low frequency update of cornering stiffness for dynamic effects, e.g. a rapid change in the tire inflation pressure or a change in the tire temperature, especially during severe driving maneuvers. Hence, even though the on-vehicle estimator provides estimates of the tire cornering stiffness taking into account the compliance effects due to the suspension and steering system and also the road roughness effects, the transient changes in the tire characteristics due to changing operating conditions of the tire are only taken into account by the adapted tire model, which takes its inputs from the tire attached sensor module, as shown in Figure 28.

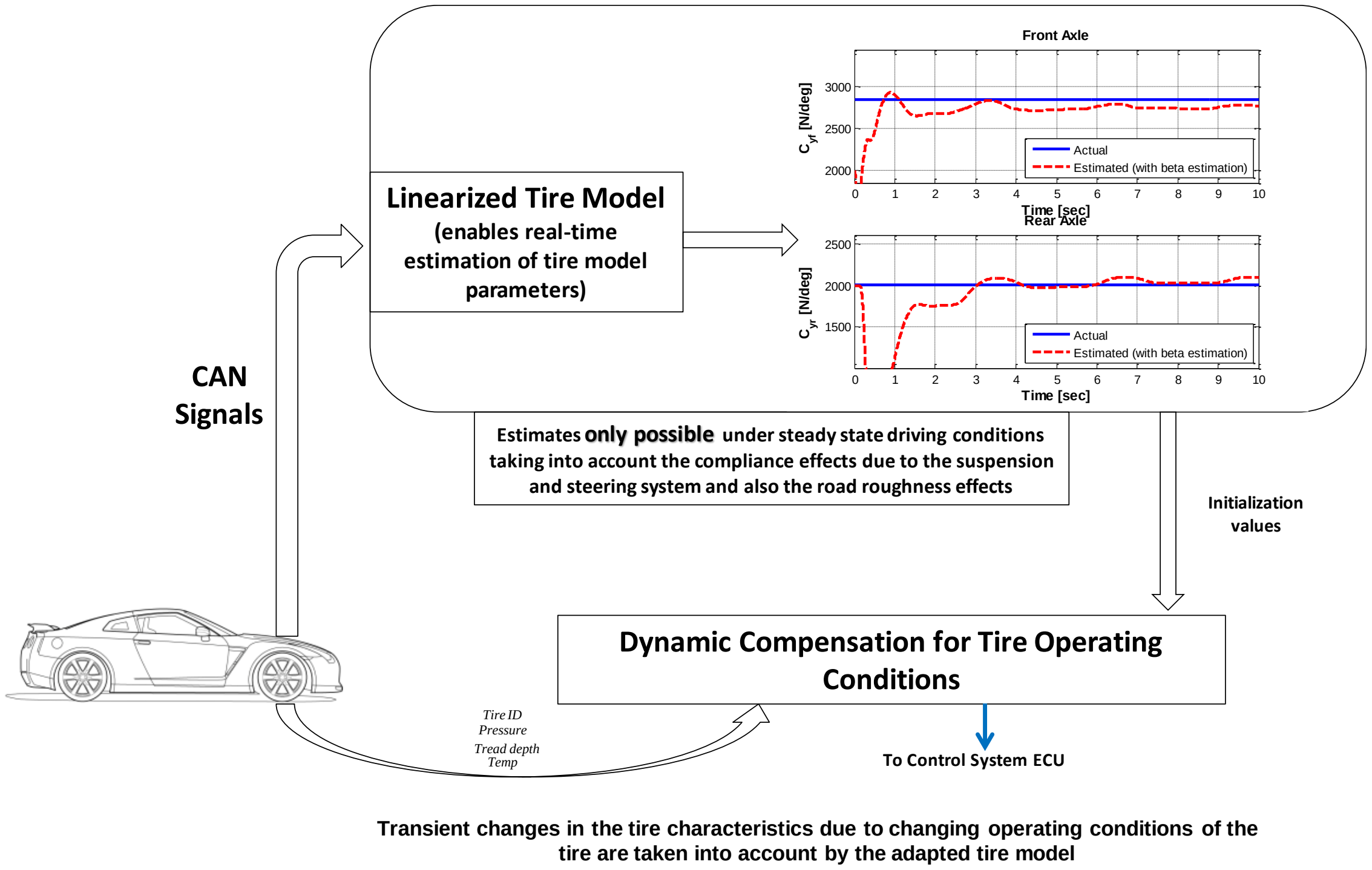

**Figure 30: On-vehicle cornering stiffness estimator**

The benefits of using an adapted tire model in a typical electronic stability control (ESC)/ electronic stability program (ESP) algorithm is presented in the next section.

## 5. Application of an adaptive tire model in an enhanced vehicle stability control system

This section investigates the possibility of enhancing the ESC performance using an adaptive tire model. The goal of an ESC system is to stabilize a vehicle in the lateral direction by controlling the yaw rate and the side slip angle. Active front/rear wheels steering [25-26] and differential wheel braking/driving [27, 28, 29] are the main actuation techniques proposed in the literature for stability control. Based on the measurements of vehicle speed, yaw rate, lateral and longitudinal accelerations, steer angle, gas pedal position and braking pressure, the ESC system derives the desired vehicle motion based on a reference vehicle model, as illustrated in Figure 31, which is compared with the actual vehicle response based on measurements, and a control action is initiated to ensure that the vehicle follows the drivers intended path without spinning out (over steering) or ploughing straight ahead (under steering).

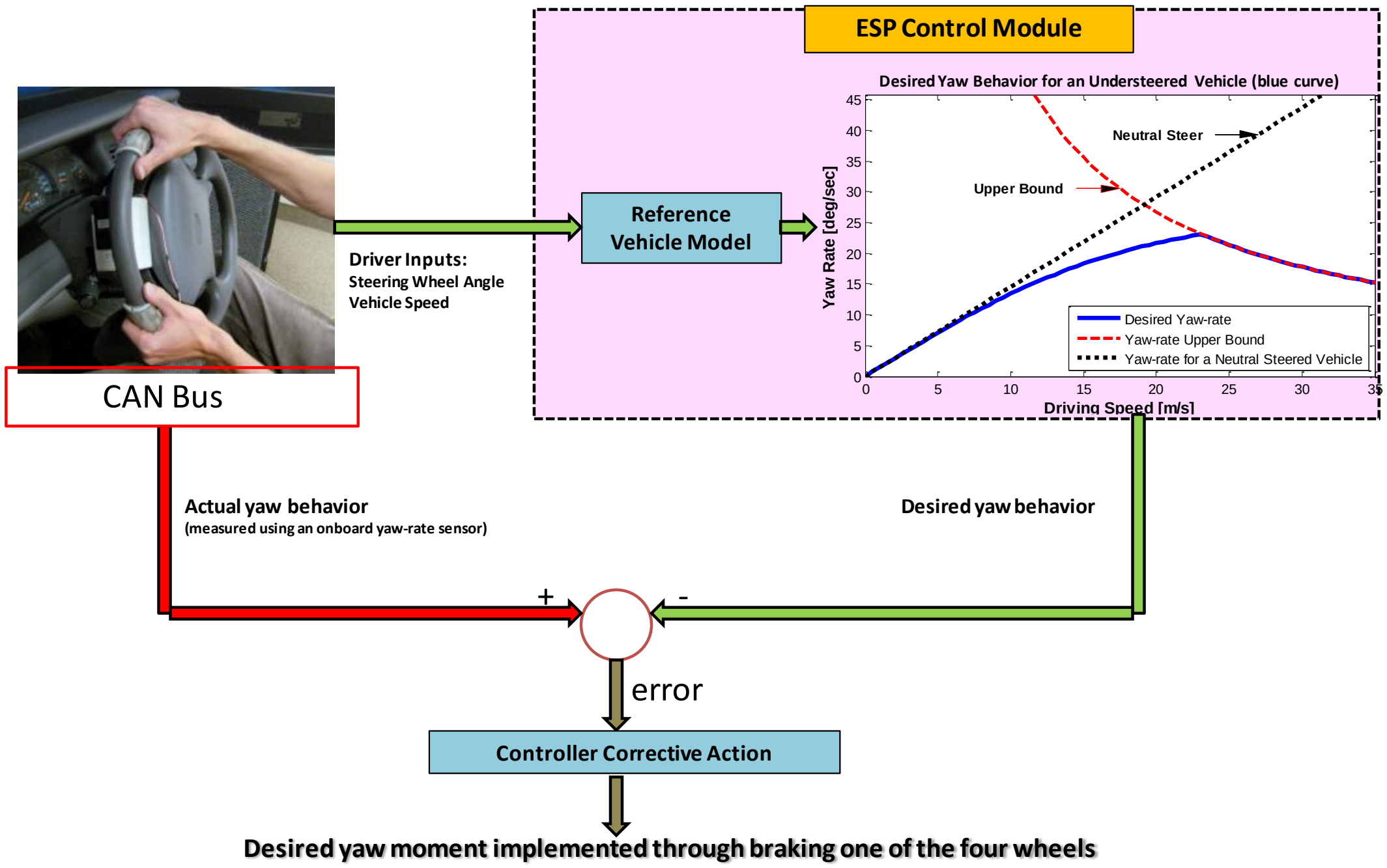

**Figure 31: ESP controller- Flow chart**

A typical ESC controller is based on hierarchical control scheme, as illustrated in Figure 32, consisting of an upper level controller and a lower level controller. The upper level controller calculates the desired yaw moment for satisfying the driver's intention.

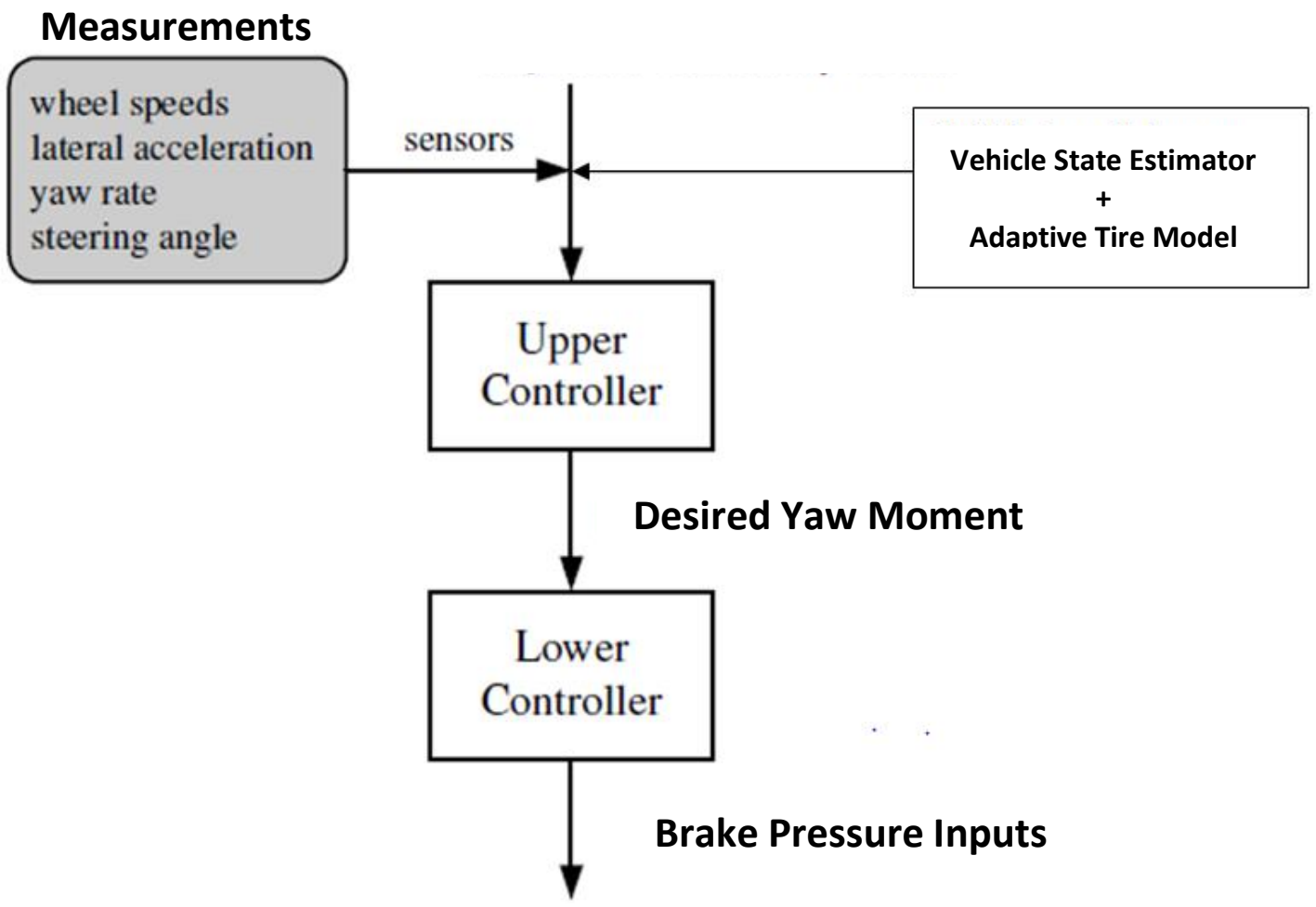

**Figure 32: Controller architecture**

The motion desired by the driver is derived from a linear bicycle model [Figure 33].

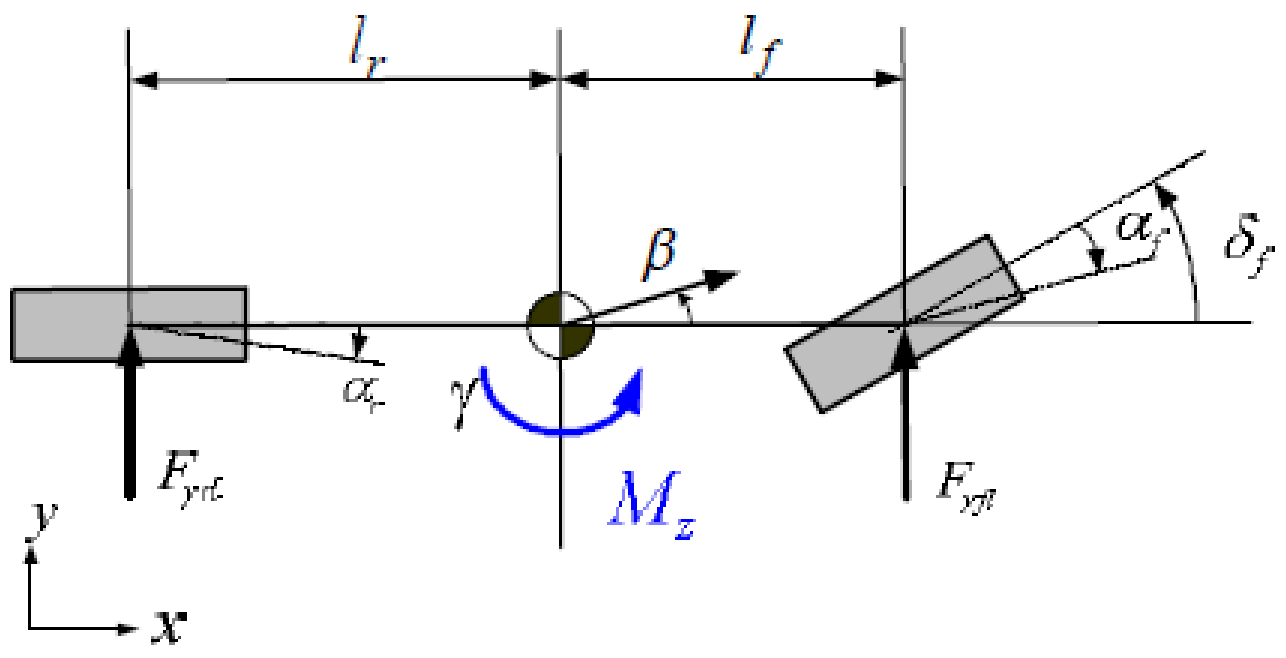

**Figure 33: Linear bicycle model vehicle model**

The desired yaw rate ($\gamma_{des}$) is calculated from the vehicle speed and the steering angle [34] and is defined as:

$$\gamma_{des} = \frac{u\delta}{(l_f + l_r)(1 + \frac{u^2}{{u_{ch}}^2})} \tag{17}$$

where the vehicle characteristic speed ($u_{ch}$) and understeer coefficient ($K_{us}$) are defined as:

$$u_{ch} = \sqrt{\frac{9.81 * (l_f + l_r)}{K_{us}}} \tag{18}$$

$$K_{us} = \left(\frac{W_f}{C_f} - \frac{W_r}{C_r}\right) \tag{19}$$

Since the lateral acceleration of the car cannot exceed the maximum coefficient of friction between the tire and the road (μ), the nominal yaw velocity must be limited to a second value by the following relation:

$$|\gamma_{des}| \le \left|\frac{\mu g}{u}\right| \tag{20}$$

This upper bound for yaw rate is to ensure that the desired yaw rate is realizable [35]. Hence, the desired yaw rate can be modified as:

$$\gamma_{des} = min\left[\frac{u\delta}{(l_f + l_r)(1 + \frac{u^2}{{u_{ch}}^2})}, \frac{\mu g}{u}\right] \tag{21}$$

Note that the desired yaw rate ($\gamma_{des}$) depends on the vehicle's understeer gradient, Kus, meaning that the vehicle understeer characteristic is preserved. Moreover, the characteristic speed ($u_{ch}$) depends mainly on the cornering stiffness of the tires. Therefore, the nominal yaw rate changes with the tire type, make and state (e.g. new or worn tire), as illustrated in Figure 34 and Figure 35.

**Front New- Rear New**

Vehicle tuned to understeer

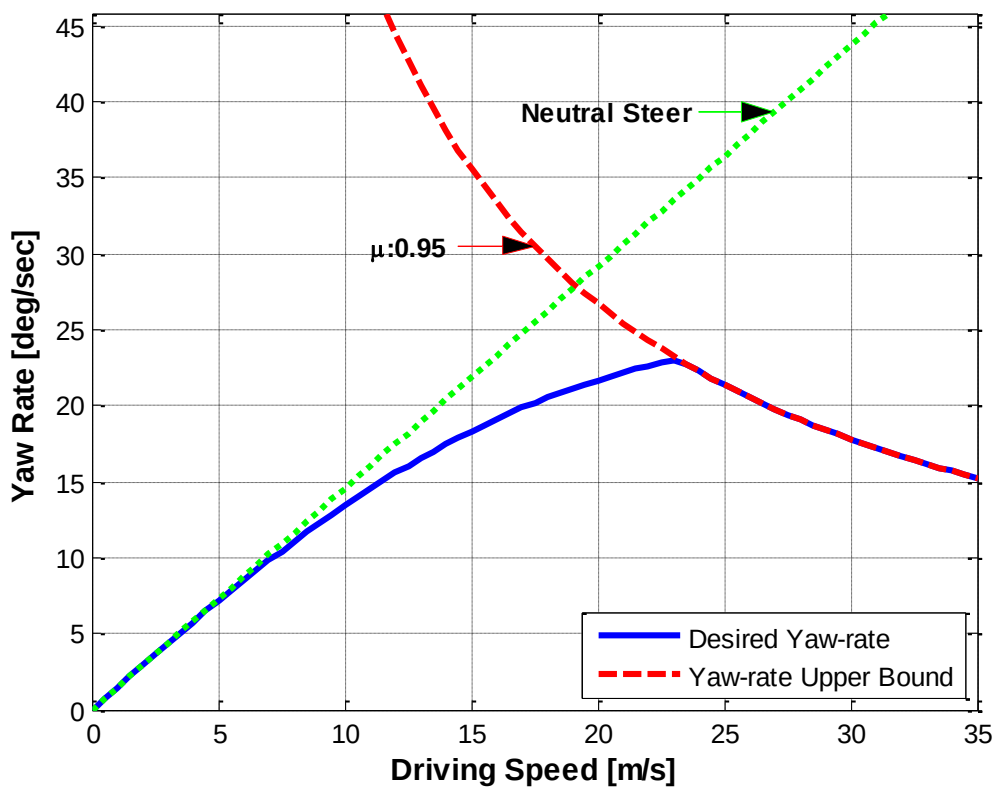

**Front New- Rear Worn (stiffer)**

Vehicle becomes <u>more understeered</u>

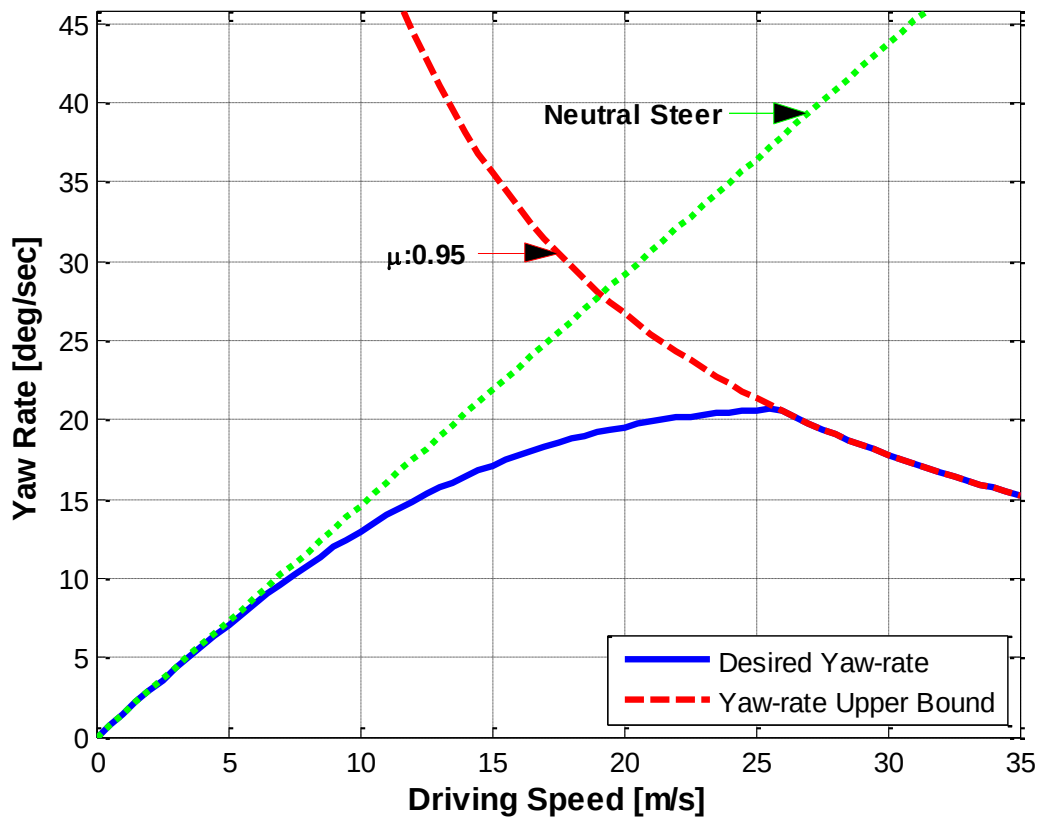

**Front Worn (stiffer)- Rear New**

Vehicle becomes <u>oversteered</u>

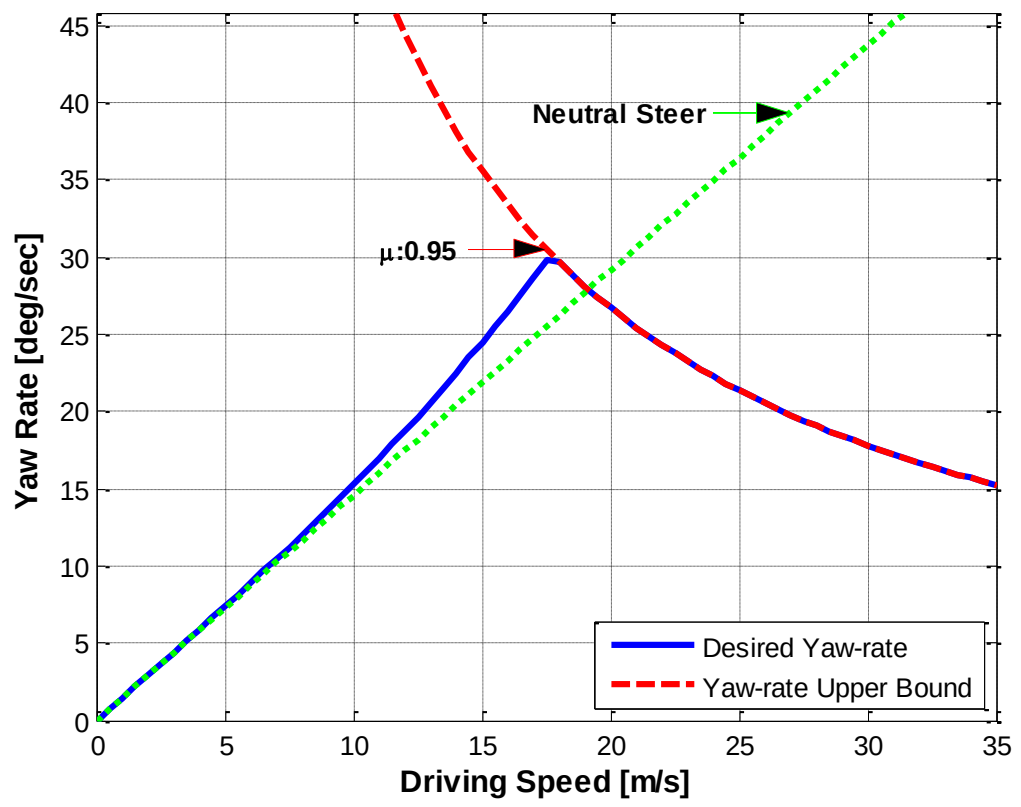

**Figure 34: Nominal yaw rate curves under different tire combinations for front and rear tires**

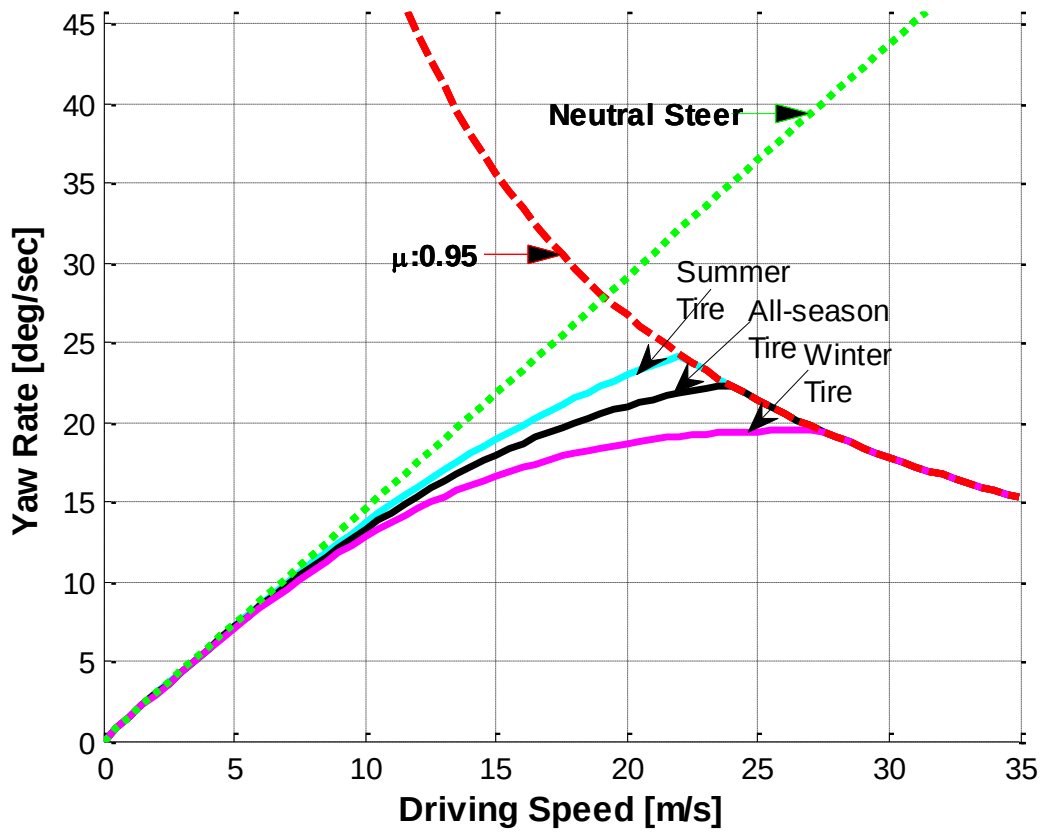

**Figure 35: Nominal yaw rate curves for different tire types (i.e. summer, all-season, winter)**

The model following control is thus sensitive to changes in the tire stiffness and ESC may suddenly change its behavior. In such cases, the vehicle behavior deviates significantly from the behavior of the linear bicycle reference model and ESC interventions can be expected for vehicle maneuvers which are well within the physical limit (i.e. brake interventions may come too early and feel harsh to the driver). Moreover, the ESC logic can worsen vehicle lateral stability by setting an improper target in dangerous cornering maneuvers. Hence, it is important for the ESC logic to account for any significant change in the tire parameters. It is proposed to use an adaptive tire model, as described in Section 3 of this paper, which can describe tire behavior with sufficient detail under different operating conditions of the tire. Unlike a traditional ESC system based on a reference vehicle model with set tire model parameters that lose effectiveness in estimating the vehicle behavior when the tire behavior changes, the enhanced ESC system proposed in this study relies only on an adaptive tire model, wherein the reference vehicle model which is used for computing the target yaw rate for ESC is updated with a new characteristic speed. For the enhanced ESC system, the desired yaw moment to satisfy the driver's intention has been determined by sliding mode control method using a 2-D bicycle model.

The sliding surface of the sliding mode controller is defined as:

$$s = \gamma - \gamma_{des} \tag{22}$$

Differentiating above equation and combining equations of lateral dynamics yields the desired yaw moment (i.e. the desired braking moment for the direct yaw control (DYC) system) as [32]:

$$M_z = \frac{I_z}{1 + \cos(\delta)} \left[ -\frac{l_f}{I_z}\left(F_{yf}\right) + \frac{l_r}{I_z}\left(F_{yr}\right) - \eta s \right] \tag{23}$$

where:
$F_{yf}$ is the front axle force
$F_{yr}$ is the rear axle force
$I_z$ is the yaw moment of inertia
$\delta$ is the road wheel angle
$l_f$ is the horizontal distance between vehicle COG and front axle
$l_r$ is the horizontal distance between vehicle COG and rear axle
$M_z$ is the desired yaw moment we desire to generate.

Here the front ($F_{yf}$) and rear ($F_{yr}$) lateral forces can be determined by the inverse of a two degree of freedom handling model given the measured lateral acceleration, yaw rate, steering angle, and estimated longitudinal tire forces [37]. In the lower level controller, the desired yaw moment needs to be implemented through braking one of the four wheels [Figure 36]. The control law is designed to select the most effective wheel to apply the brake torque according to the following situation.

- **Understeer condition**: In this case, the absolute value of the vehicle yaw rate is always smaller than the absolute value of the desired vehicle yaw rate. The system applies the brake on the inner rear wheel to give an inward compensating moment. This moment suppresses the under steering behavior and prevents the car from ploughing out of the curve.

- **Oversteer condition**: In this condition, the absolute value of the vehicle yaw rate is always greater than the absolute value of desired vehicle yaw rate. The system applies the brake on the outer wheel(s) to give an outward compensating moment. This moment prevents the car from rotating in to the curve and avoids rear wheel side slip.

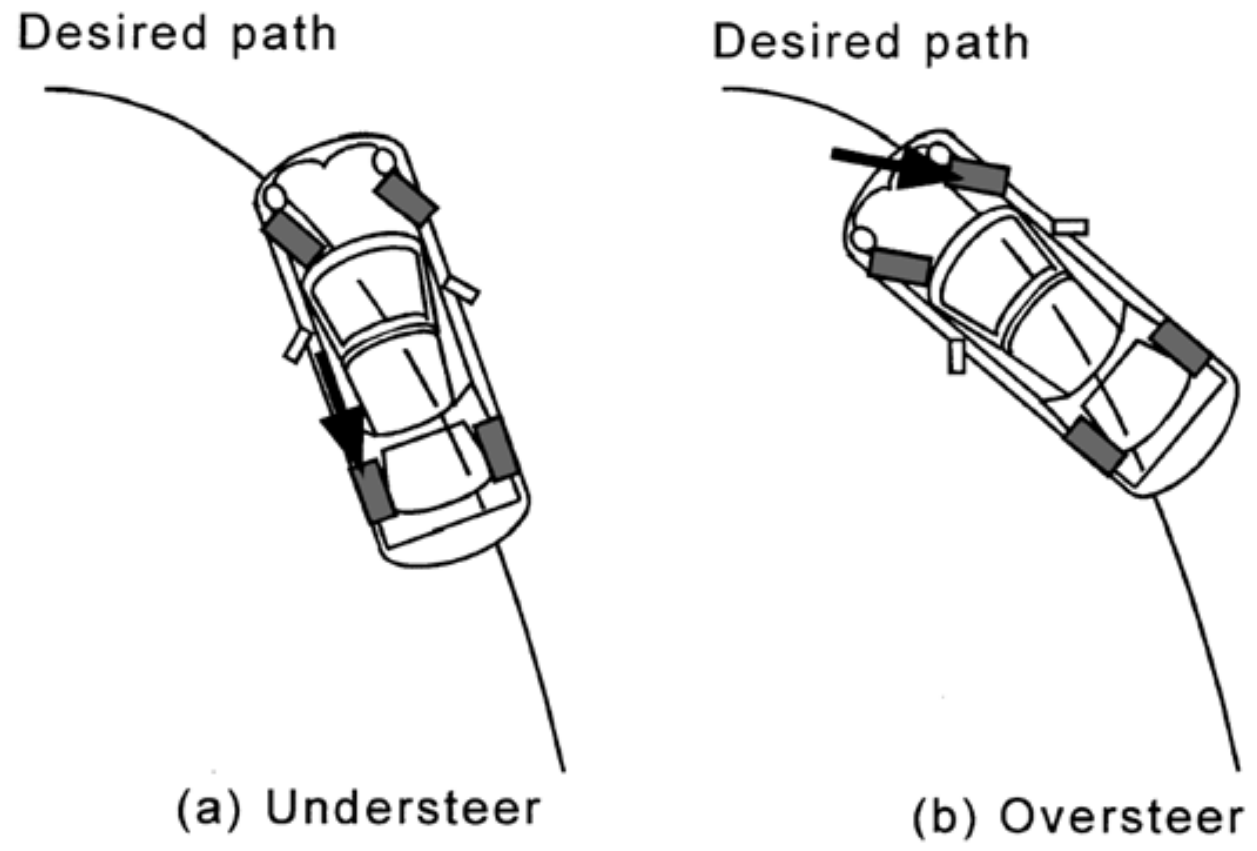

**Figure 36: Location of controlled braking torque for the yaw rate controller in response to steering input. Shown here is the configuration for a left-hand turn [36]**

Based on the above analysis and assuming counterclockwise positive, the lower-level controller law is described by the decision rules shown in Table 4.

**Table 4: Brake rules for the DYC implementation [33]**

| Desired yaw rate $r_d$ | Error $r - r_d$ | Desired Brake Moment $Mz$ | Braked wheel |
|---|---|---|---|
| + | + | + | None |
| + | + | − | Front left |
| + | − | + | Rear right |
| + | − | − | None |
| − | + | + | None |
| − | + | − | Rear left |
| − | − | + | Front right |
| − | − | − | None |

Finding which wheel to brake, the required brake torque is simply calculated by using:

$$T_b = \frac{M_z R_w}{t_r/2} \tag{24}$$

where *Rw* is the wheel radius and *tr* represents the track width of the vehicle. The braking action is achieved with an ABS controller to ensure a reasonable longitudinal slip. The controller adjusts longitudinal slip targets for the ABS controller to appropriate targets when slip angle is present using a simple model of combined slip. Combined slip model provides a friction limitation in the lower controller. This makes the tire force "commanded" by the controller always within the peak force potential, i.e. friction circle, of the tire.

The following open loop (i.e. without driver's feedback) maneuver has been considered for simulation:

- 0.7 Hz sine with dwell test. This transient maneuver is considered by the National Highway Traffic Safety Administration (NHTSA) for electronic stability control evaluation [30-31], since it best excites an oversteer response from the vehicle. The front steering angle course issued during the maneuver is shown in Figure 37.

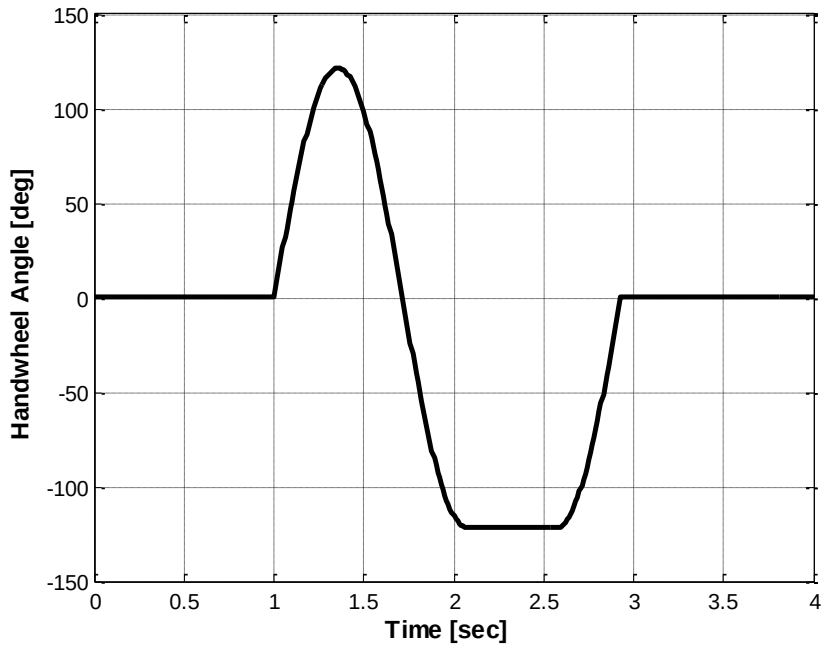

**Figure 37: Front steering angle input for the 0.7 Hz sine with dwell maneuver**

- The sine with dwell maneuver is performed at a starting speed of about 80 km/h (50 mph), with increasing hand wheel values until excessive oversteer occurred (i.e. the vehicle direction 4 s after the completion of the steer input is more than 90◦ from the initial path, see [30]) or a hand wheel angle amplitude of 330◦ was reached

Simulation results shown in Figure 38 demonstrate the effectiveness of a traditional ESC system in its ability to stabilize the vehicle by minimizing the vehicle body sideslip angle. The maximum hand wheel angle amplitude was restricted to 120 degrees.

(a) Vehicle without an ESC system

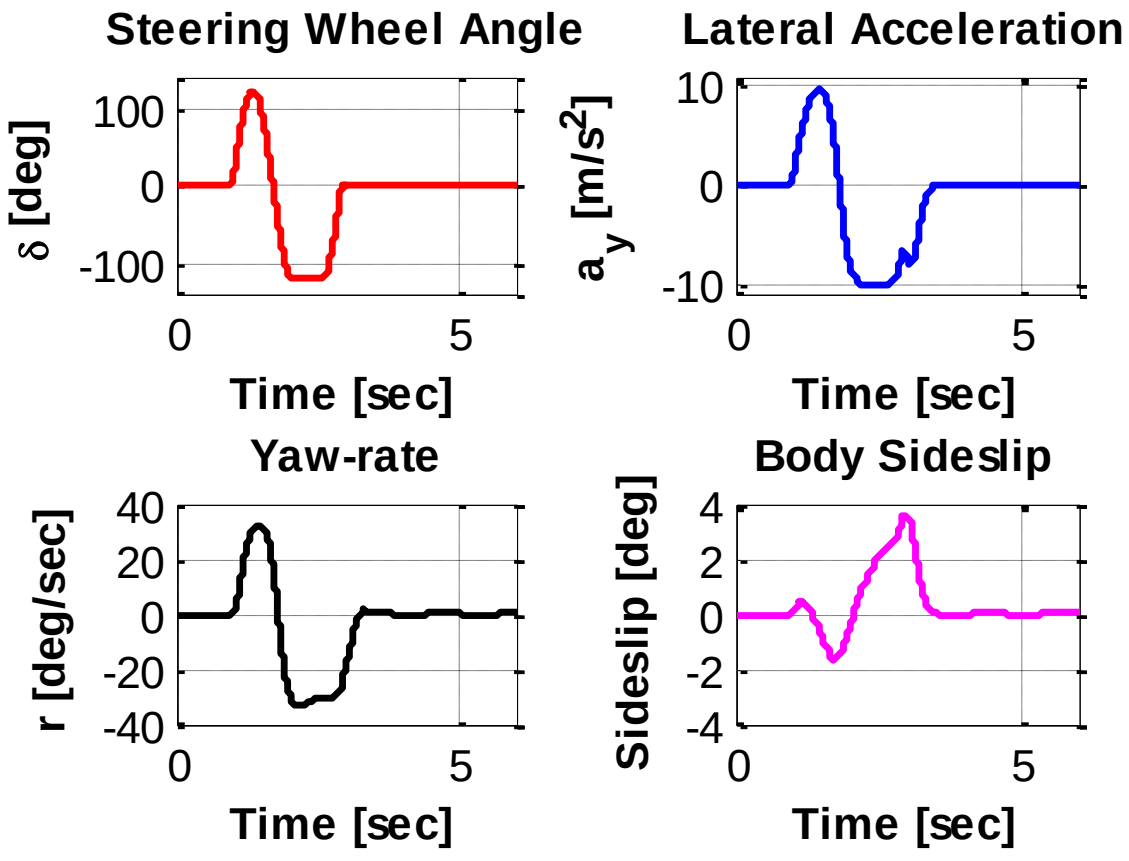

(b) Vehicle with an ESC system

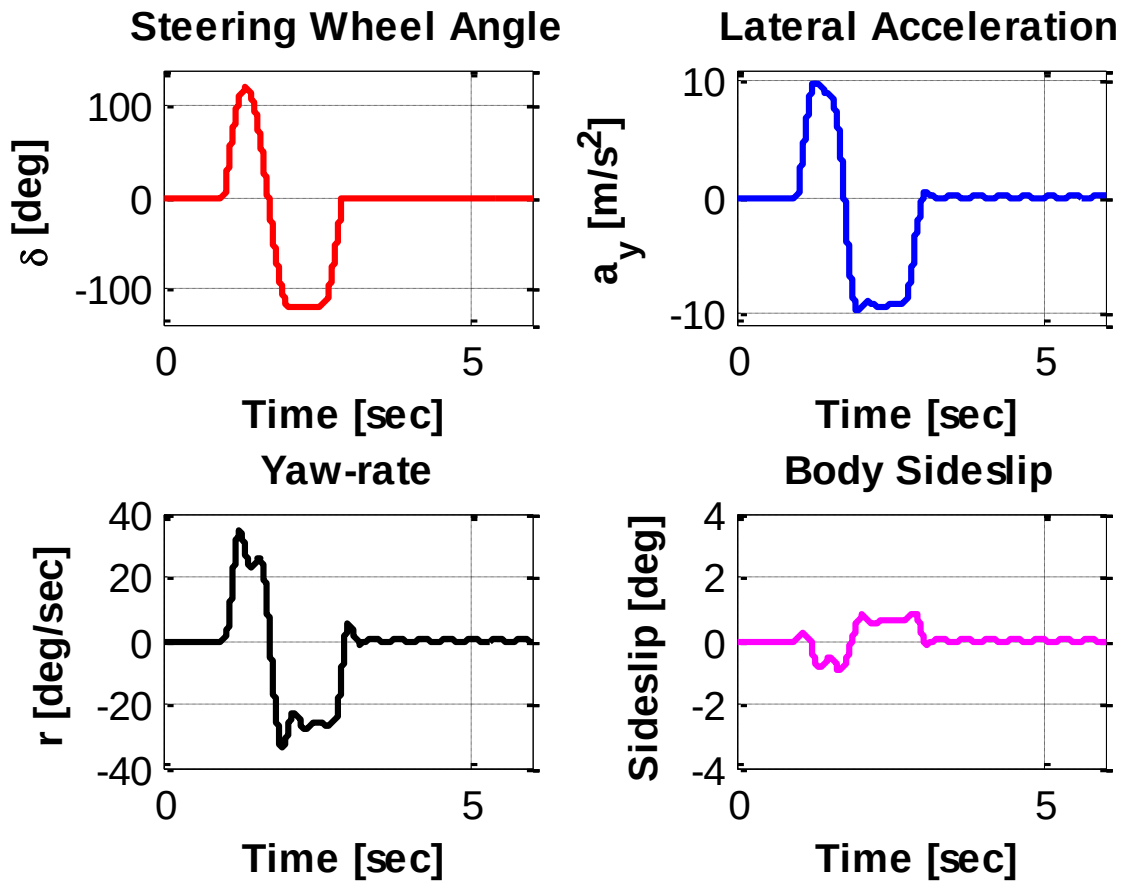

(c) Vehicle body sideslip comparision

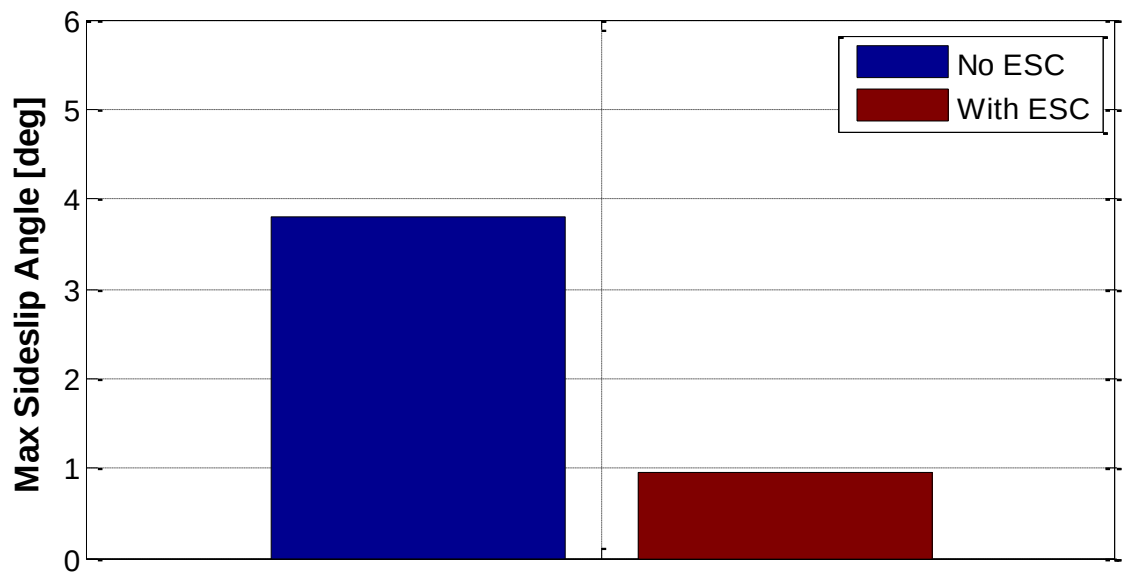

**Figure 38: Vehicle behavior with and without an ESC system**

A further increase in the hand wheel angle amplitude results in the loss of lateral stability for the vehicle not equipped with an ESC system (Figure 39a). On the contrary, the vehicle with an ESC system doesn't lose lateral stability (Figure 39b).

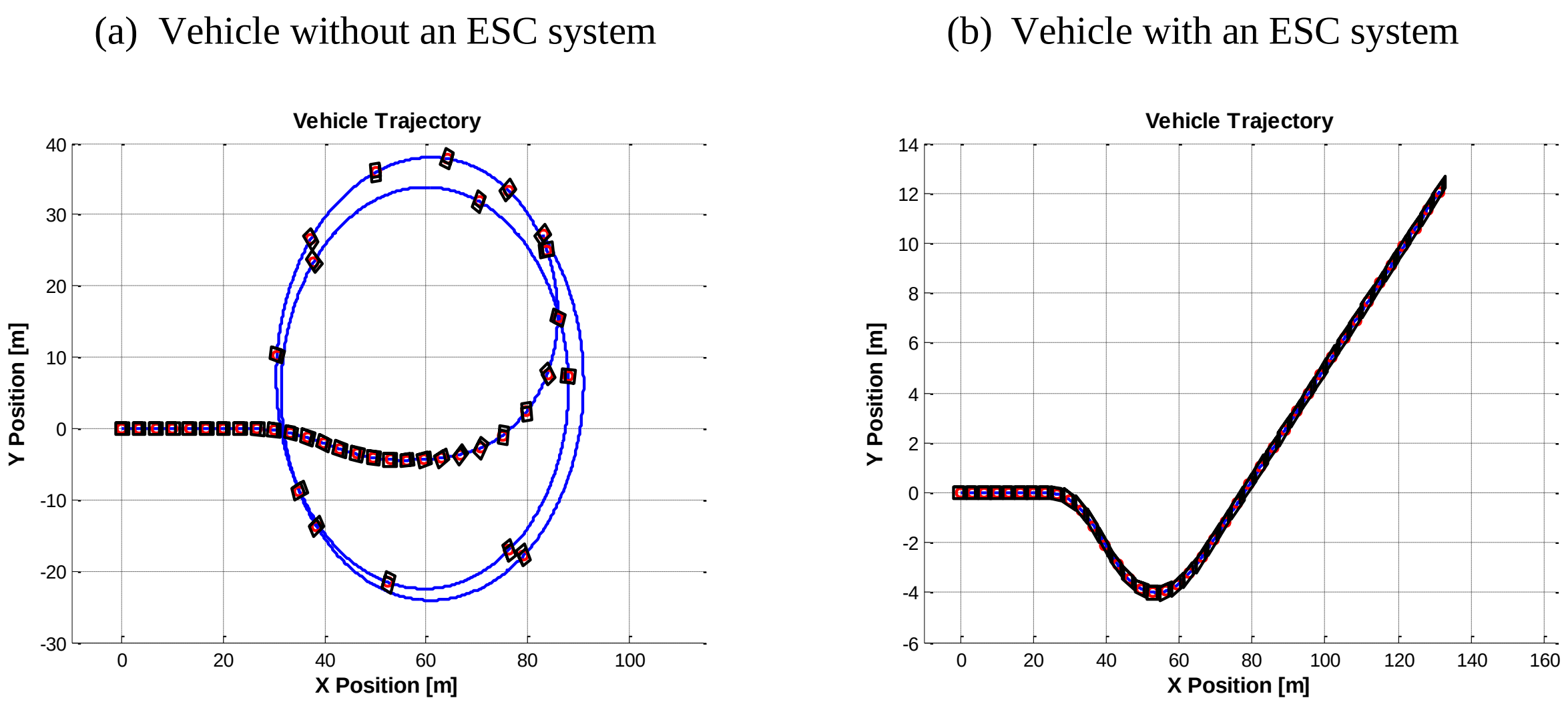

**Figure 39: Vehicle trajectory comparison**

Simulation results shown in Figure 40 demonstrate the potential benefits of an enhanced ESC system using an adaptive tire model in comparison to a traditional ESC system using a tire model with fixed model parameters.

The vehicle with an adaptive tire model can follow the yaw rate more accurately than the vehicle using a non-adaptive reference controller model (Figure 40a). Moreover, using an adaptive tire model results in better lateral stability of the vehicle, i.e. lower body sideslip angle, as shown in Figure 40b.

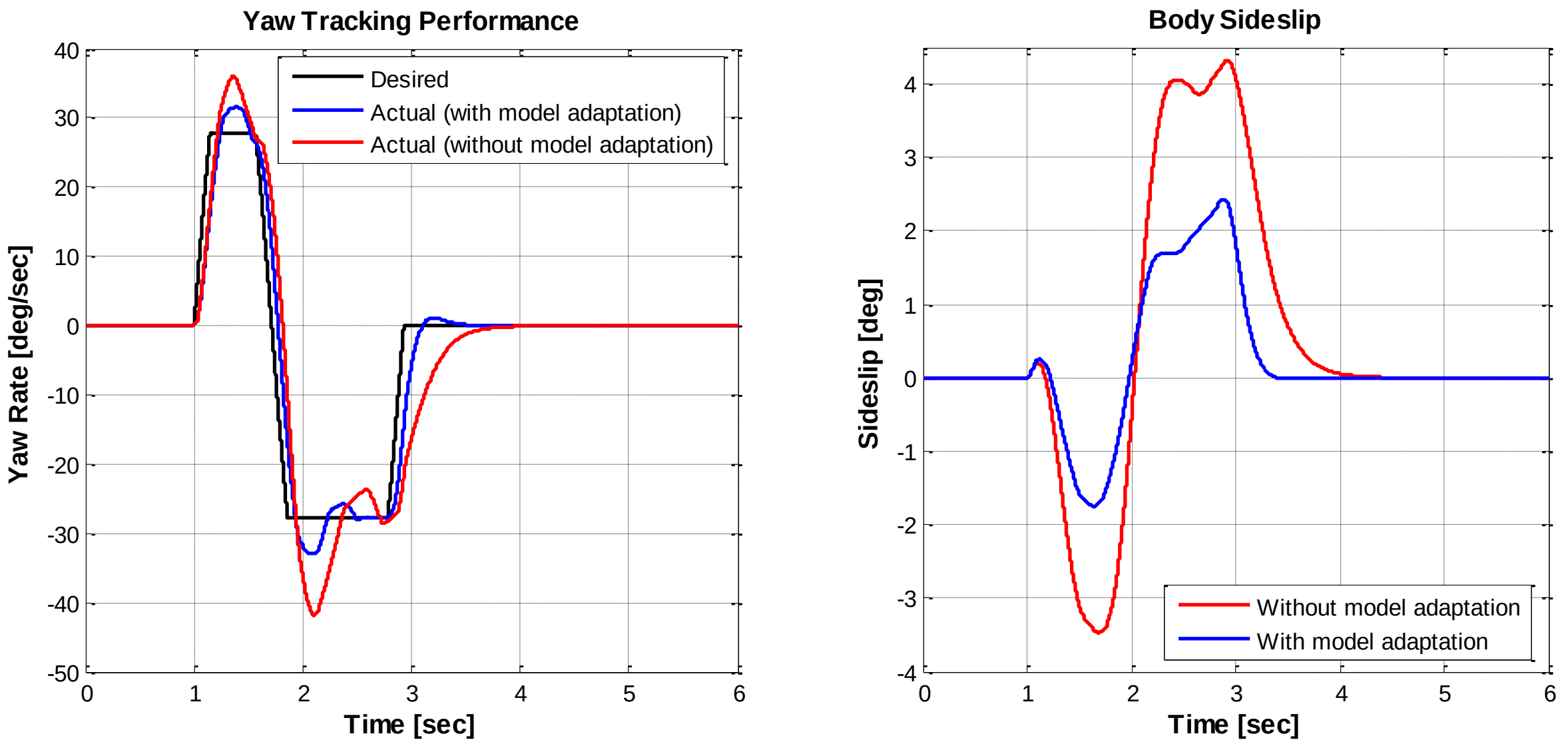

**Figure 40: Performance comparison- Enhanced ESC with model adaptation versus traditional ESC without model adaptation; (a) yaw rate tracking performance, (b) vehicle body sideslip angle**

Simulations results shown in Figure 41 indicate that the adaptive tire model based ESC algorithm imposes a lower work-load on the control signals (i.e. lesser activity seen in Figure 41b in comparison to Figure 41a), which results in lower levels of interference with the driver inputs. Moreover, more ESC interventions result in a larger drop in the vehicle speed (Figure 41a), which is typically undesirable.

(a) Without model adaptation (b) With model adaptation

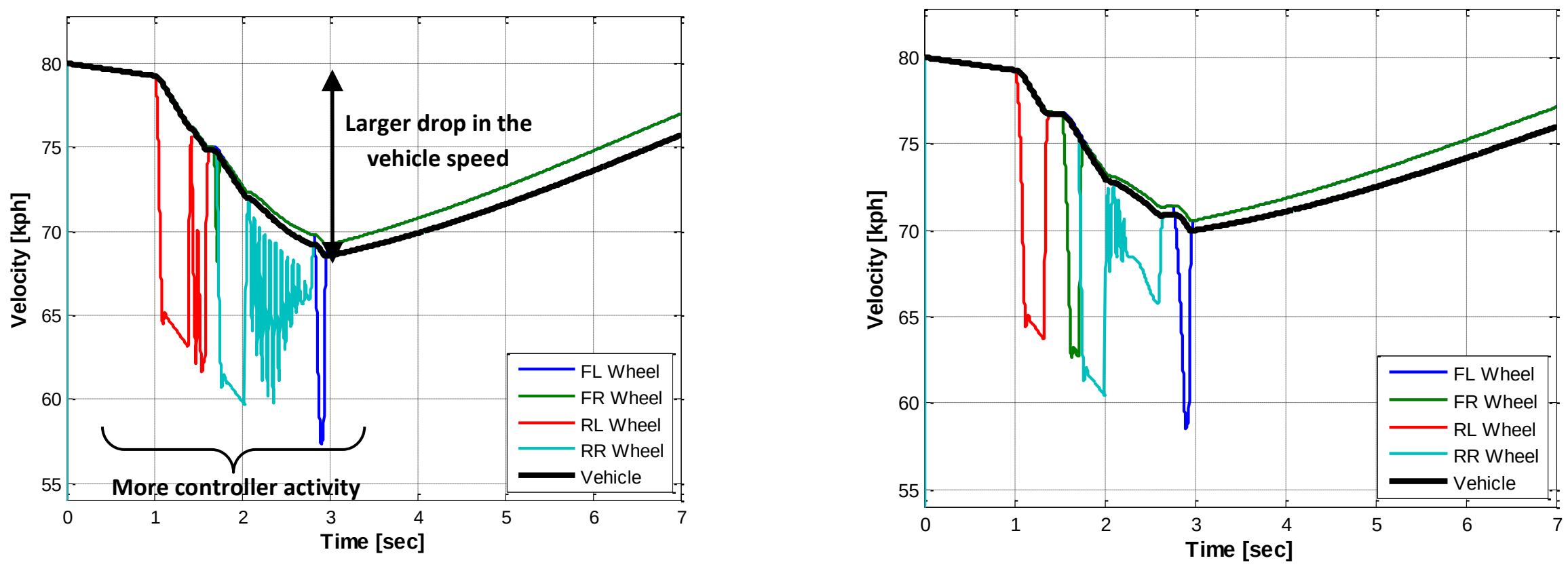

**Figure 41: Comparison of the controller activity level**

Hence, results presented in this section of the paper demonstrate the performance and effectiveness of an enhanced ESC control system based on an adaptive tire model.

The usage of an adaptive tire model in combination with a modified ABS controller which utilizes this information to dynamically adapt its tuning parameters and therefore maximizes the available force utilization from the tire is expected to improve the ABS performance. Consequently, it will decrease stopping distance. The high-level operation of such a controller is shown in Figure 42 where inputs such as inflation pressure, temperature, tread depth are sent to the adaptive magic formula which uses the construction information of the tire (summer, winter and all-season) to determine the appropriate scaling factors. The final magic formula coefficients are then used to compute relevant tire characteristics such as braking stiffness, optimal slip ratio and peak grip respectively which would then be sent as inputs to the modified ABS controller. The investigation of intelligent tires that would provide tire data to such adaptive-ABS algorithms is being undertaken as part of future work.

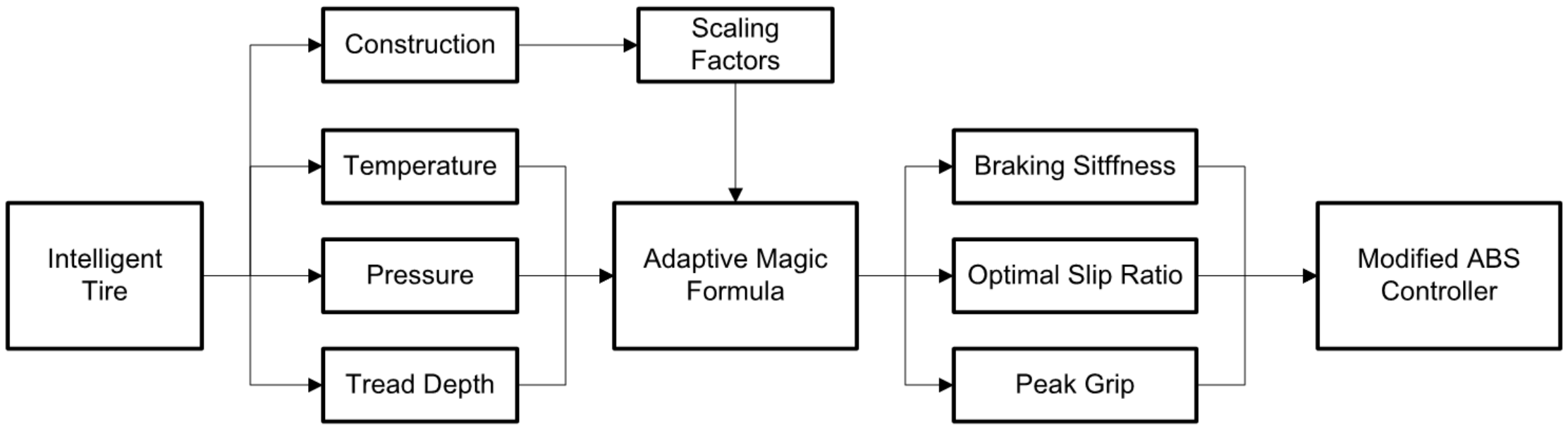

**Figure 42 - Modified ABS Controller Operation**

## 6. Conclusion

This paper presents a state-of-the-art review of the developments that have been made to extend tire models to improve their prediction capabilities under various operating conditions of the tire. Thereafter, details of an adaptive magic formula (MF) tire model capable of coping with changes to the tire operating conditions are presented. More specifically, extensions have been made to the magic formula expressions for tire cornering stiffness and peak grip level, to account for variations in the tire inflation pressure, load, tread-depth and temperature. As a next step, the benefits of using an adaptive tire model for vehicle control system applications is demonstrated through simulation studies for an enhanced ESC system using an adaptive tire model in comparison to a traditional ESC system based on a fixed reference model. The ESC system used in this study is a model based controller based on a differential braking control strategy using yaw rate feedback. Comparative analysis performed with the NHTSA sine with dwell maneuver confirms that an enhanced ESC controller using an adaptive tire model results in better vehicle stabilization and improved yaw-rate tracking. Therefore, using an adaptive tire model ensures that the controller exhibits robustness to meet the specified requirements even under varying operating conditions of the tire. Although the use of tire data for ESC controller indicates potential improvements in performance, sufficient caution must be practiced in their commercial implementation to ensure robust tire identification and data transmission with adequate fallback options.

### Note

The views and opinions expressed in this article are those of the authors and do not necessarily represent the views of the The Goodyear Tire & Rubber Company.

### References

[1] H.B. Pacejka, Tyre and Vehicle Dynamics, Elsevier, Oxford, U.K, 2005
[2] I.B.A.o.h. Veld, Enhancing the MF-Swift Tyre Model for Inflation Pressure Changes, in: Department of Mechanical Engineering, Dynamics and Control Group, Eindhoven University of Technology, Eindhoven, 2007
[3] Février P., Martin H., Fandard G, Method for simulating the thermomechanical behaviour of a tyre rolling on the ground. Patent from Michelin. {March} 2008, international ref. number: WO 2008/025892 A1
[4] Février P., Thermal and Mechanical Tire Force & Moment Model presentation, 4th IPG Technology Conference, Ettlingen, September 23-24, 2008
[5] SÉBASTIEN CHALIGNÉ, Tire Thermal Analysis And Modeling, Master’s Thesis in Automotive Engineering, CHALMERS UNIVERSITY OF TECHNOLOGY, Göteborg, Sweden 2011
[6] Masahiko M., Development of Tire Side Force Model Based on “Magic Formula” with the Influence of Tire Surface Temperature, R&D Review of Toyota CRDL, Vol. 38, No. 4, October 2003
[7] Sorniotti, A. and Velardocchia, M., "Enhanced Tire Brush Model for Vehicle Dynamics Simulation," SAE Technical Paper 2008-01-0595, 2008, doi:10.4271/2008-01-0595
[8] K. Guo, Y. Zhuang, D. Lu, S.-k. Chen and W. Lin, A study on speed-dependent tyre–road friction and its effect on the force and the moment, Vehicle System Dynamics. 43 (2005), pp. 329-340
[9] Kart Handling Theoretical Basics: Available online at: http://www.gokarthandling.com/files/80/File/PDF_111-119_lock.pdf
[10] Yong Li, Shuguang Zuo, Lei Lei, Xianwu Yang, and Xudong Wu, Characteristics analysis of lateral vibration of tire tread Journal of Vibration and Control December 2011 17

[11] B N J Persson, Rubber friction and tire dynamics, J. Phys.: Condens. Matter 23 (2011) 015003 (14pp)
[12] Peter Zegelaar, PhD Thesis, “The dynamic response of tyres to brake torque variations and road unevennesses” Available Online: http://www.tno.nl/downloads%5CDT_PhD_Thesis_Zegelaar2.pdf
[13] Hermann Winner, Marcus Reul, Enhanced Braking Performance by Integrated ABS and Semi-Active Damping Control, TU Darmstadt, Chair Of Automotive Engineering, Germany, 2009
[14] Ari J Tuononen , Laser triangulation to measure the carcass deflections of a rolling tire, Meas. Sci. Technol. 22 (2011) 125304 (8pp)
[15] Sorniotti, Aldo. "Tire Thermal Model for Enhanced Vehicle Dynamics Simulation." Training 2005: 12-15
[16] Kelly, D. P., and R. S. Sharp. "Time-optimal control of the race car: influence of a thermodynamic tyre model." Vehicle System Dynamics 50.4 (2012): 641-662
[17] Fevrier, P., and O. Le Maitre. "Tire temperature modelling: Application to race tires." proceedings VDI 13th Int. Congress numerical analysis and simulation in vehicle engineering. 2006
[18] Netsch, L., et al. "T 3 M—TÜV Tire temperature method—a breakthrough methodology for evaluating tire robustness, performance and wear." Proceedings of 31st FISITA World Automotive Congress. 2006
[19] H. B. Pacejka, "Tyre brush model," in Tyre and Vehicle Dynamics. vol. 2nd ed ed Oxford, U.K: Elsevier, 2005, pp. 93-134
[20] Available online at: www.vtt.fi/apollo
[21] Available online at: http://friction.vtt.fi
[22] Available online at: http://www.bridgestone.eu/press/press-releases/2011/bridgestone-announces-new-tyre-technology-for-determining-road-surface-conditions-based-on-the-concept-of-cais
[23] Available online at: http://www.motorauthority.com/news/1044745_pirelli-schrader-team-up-for-computerized-cyber-tire
[24] Available online at: http://www.conti-online.com/generator/www/de/en/continental/automotive/themes/commercial_vehicles/chassis/chassis_control/intelligent_tire_system/intelligent_tire_system_en.html
[25] J. Ackermann, T. Bunte, and D. Odenthal, “Advantages of active steering for vehicle dynamics control,” 1999
[26] J. Ackermann, “Robust yaw damping of cars with front and rear wheel steering,” in Proceedings of the 31st Conference on Decision and Control, Tucson-Arizona, Dec. 1992, pp. 2586–2590
[27] R. Daily and D. Bevly, “The use of GPS for vehicle stability control systems,” IEEE Transactions on Industrial Electronics, vol. 51, no. 2, pp. 270–277, 2004
[28] S. Anwar, “Predictive yaw stability control of a brake-by-wire equipped vehicle via eddy current braking,” in American Control Conference, 2007, 9-13 July 2007, pp. 2308–2313
[29] A. T. van Zanten, R. Erthadt, and G. Pfaff, “VDC, the vehicle dynamics control of Bosch,” in International Congress and Exposition, 1995. Proceeding of the, ser. SAE950759, March 1995, pp. 9–26
[30] G.J. Forkenbrock, D. Elsasser, and B. O’Hara, NHTSA’s Light Vehicle Handling and ESC Effectiveness Research
Program, ESV Paper Number 05-0221 (2005)
[31] G.J. Forkenbrock and P. Boyd, Light Vehicle ESC Performance Test Development, ESV Paper Number 07-0456 (2007)
[32] Available online at: http://www.ocf.berkeley.edu/~stephenc/pdf/tvsc.pdf
[33] Ding, Nenggen and Taheri, Saied(2010) 'An adaptive integrated algorithm for active front steering and direct yaw moment control based on direct Lyapunov method', Vehicle System Dynamics, 48: 10, 1193 — 1213
[34] J.Y. Wong, Theory of Ground Vehicles, Third Edition, John Wiley & Sons, 2001
[35] Piyabongkarn, D.; Rajamani, R.; Grogg, J.A; Lew, J.Y., "Development and Experimental Evaluation of a Slip Angle Estimator for Vehicle Stability Control," Control Systems Technology, IEEE Transactions on , vol.17, no.1, pp.78,88, Jan. 2009

[36] Taehyun Shim & Donald Margolis (2001) Using μ Feedforward for Vehicle Stability Enhancement, Vehicle System Dynamics: International Journal of Vehicle Mechanics and Mobility, 35:2, 103-119
[37] Sill, J.H.; Ayalew, B., "Managing axle saturation for vehicle stability control with independent wheel drives," American Control Conference (ACC), 2011 , vol., no., pp.3960,3965, June 29 2011-July 1 2011
[38] Sierra, C., Tseng, E., Jain, A. and Peng, H., 2006. Cornering stiffness estimation based on vehicle lateral dynamics. Vehicle System Dynamics, 44(sup1), pp.24-38.